# Direct simulation of second sound in graphene by solving the phonon Boltzmann equation via a multiscale scheme

Xiao-Ping Luo[a*], Yang-Yu Guo[b*], Mo-Ran Wang[c] and Hong-Liang Yi[a†]

a *Key Laboratory of Aerospace Thermophysics, Ministry of Industry and Information Technology, School of Energy Science and Engineering, Harbin Institute of Technology, 92 West Dazhi Street, Harbin 150001, China*

b *Institute of Industrial Science, The University of Tokyo, Tokyo 153-8505, Japan*

c *Key Laboratory for Thermal Science and Power Engineering of Ministry of Education, Department of Engineering Mechanics and CNMM, Tsinghua University, Beijing 100084, China*

## Abstract

The direct simulation of the dynamics of second sound in graphitic materials remains a challenging task due to lack of methodology for solving the phonon Boltzmann equation in such a stiff hydrodynamic regime. In this work, we aim to tackle this challenge by developing a multiscale numerical scheme for the transient phonon Boltzmann equation under Callaway's dual relaxation model which captures well the collective phonon kinetics. Comparing to traditional numerical methods, the present multiscale scheme is efficient, accurate and stable in all transport regimes attributed to avoiding the use of time and spatial steps smaller than the relaxation time and mean free path of phonons. The formation, propagation and composition of ballistic pulses and second sound in graphene ribbon in two classical paradigms for experimental detection are investigated via the multiscale scheme. The second sound is declared to be mainly contributed by ZA phonon modes, whereas the ballistic pulses are mainly contributed by LA and TA phonon modes. The influence of temperature, isotope abundance and ribbon size on the second sound propagation is also explored. The speed of second sound in the observation window is found to be at most 20 percentages smaller than the theoretical value in hydrodynamic limit due to the finite umklapp, isotope and edge resistive scattering. The present study will contribute to not only the solution methodology of phonon Boltzmann equation, but also the physics of transient hydrodynamic phonon transport as guidance for future experimental detection.

[*] These authors contributed equally to this work.

[†] Corresponding author. E-mail address: yihongliang@hit.edu.cn

## 1. Introduction

With the rapid development of microelectromechanical systems and nanotechnology in recent decades, thermal management of micro- and nano-electronics has become a challenging bottleneck that limits the further technology development in the semiconductor industry [1-6]. It becomes an essential task to seek materials with high thermal conductivity for the dissipation of heat from nanoscale hot spots. For semiconductor materials in solid-state electronics, heat transport is mainly mediated by phonons, quantized lattice vibrations. The thermal conductivity of most semiconductor materials is limited by the dominant momentum-destroying phonon scattering (resistive process) causing transport resistance [7]. Therefore, a natural way to achieve high thermal conductivity is to find materials with dominant momentum-conserving phonon scattering (normal process), where hydrodynamic phonon transport becomes the main heat transport mechanism [8,9].

There have been intensive theoretical and experimental studies on phonon hydrodynamics around the last mid-century due to the interest in exploration of heat waves (or called second sound) in cryogenic condensed systems [1,10-14]. Yet the window condition of hydrodynamic phonon transport is usually very narrow and at extremely low temperature (around 10K or lower) in three-dimensional materials [11-14], which limits its practical application in thermal management often at ordinary temperature. In recent years, there are renewed interests in hydrodynamic phonon transport due to the first-principle theoretical prediction of its strong effect in graphitic materials even close to room temperature [15-18]. The theoretical prediction has been confirmed in a recent experimental report [19] where second sound is observed in graphite at temperatures up to 100 K using time-resolved optical techniques. In terms of the theoretical study, the window conditions of the two main phonon hydrodynamic phenomena: the steady-state phonon Poiseuille flow and transient second sound, have been well formulated for the graphene [16], other two-dimensional materials [15,20] and graphite [17]. The dispersion relation of second sound in single-wall carbon nanotube has been derived in the limit of only normal process (hydrodynamic limit) [21]. Furthermore, the detailed dynamics (including temperature and heat flux profiles) and thermal transport resistance of both in-plane (*i.e.* phonon Poiseuille flow) [18,22] and cross-plane [23,24] steady-state hydrodynamic phonon transport in graphene ribbon have been well studied. In strong contrast, it remains to investigate the transient dynamics (including the evolution of temperature and heat flux profiles) of the second sound propagation in graphitic materials, which is the main aim of the present work. The direct simulation of second sound is crucial for both practical application in heat dissipation of high-frequency nano-electronics and guidance for future experimental detection of second sound in graphene ribbon. The theoretical modeling of hydrodynamic phonon transport is usually based on the phonon Boltzmann equation (PBE). In general, there are two kinds of treatment for the scattering term in PBE: (*i*) the *ab initio* full scattering term, (*ii*) relaxation time approximation. The numerical solution of PBE with the *ab initio* full scattering term has been achieved for both the homogeneous heat transport in infinite graphene [15,16] and the steady-state heat transport in graphene ribbon [18,25]. As the numerical solution of PBE with the *ab initio* full scattering term is computationally expensive even for steady-state cases, relaxation time approximations are often adopted instead. The single mode relaxation time (SMRT) approximation is known to much underestimate the thermal conductivity of graphene due to the collective effect of phonon normal process [15,26]. In contrast, the Callaway's dual relaxation model [27] represents a better approximation to the full scattering term of PBE and is widely used in modeling and analyzing hydrodynamic phonon transport in graphitic materials [17,18,21-24,28,29]. For transient heat transport cases, the solution of PBE with the *ab initio* full scattering term is extremely challenging and is not available in the literature to the author's best knowledge. Therefore, the simulation of second sound will be conducted based on a numerical solution of PBE under Callaway's dual relaxation model in the present work.

The direct numerical solution of PBE under Callaway's dual relaxation model remains still in its infancy stage, especially for transient situations. A gray lattice-Boltzmann model was developed for the Callaway's model based on a Chapman-Enskog expansion [30]. It remains a difficult task to include the spectral phonon properties into the lattice Boltzmann model. A semi-analytical solution of the PBE under Callaway's model has been obtained by using the integral formulation for steady-state hydrodynamic phonon transport in simple geometrical structures [17,24]. Yet it is not realistic to obtain analytical solutions of the PBE as an integro-differential equation for more general geometries. Some important progress has been made in recent work where a discrete-ordinate method [22] and a Monte Carlo scheme [23] have been developed for direct numerical solutions of the steady-state PBE under Callaway's dual relaxation model. However, there still lacks a robust and efficient approach to numerical solution of transient PBE under Callaway's dual relaxation model, which is another aim of the present work. From the perspective of numerical computation, deterministic grid-based methods (such as discrete-ordinate method, finite volume method, finite element method, etc.) for PBE under SMRT approximation can be extended for the Callaway's model in a straightforward way. Unfortunately, these methods will encounter numerical difficulties when the phonon mean free path (MFP) or relaxation time is very small comparing to the system characteristic length or time, where the PBE becomes stiff. This situation is very relevant to the simulation of second sound, the window condition of which requires the relaxation time of phonon normal process much smaller than the external heat pulse duration. As a result, these traditional deterministic grid-based methods may suffer possible order reduction, numerical instability and serious efficiency loss in such a stiff regime. Recently, a novel multiscale method, named discrete unified gas kinetic scheme (DUGKS) has been proposed to model gas transport problems [31,32]. Numerical tests have demonstrated that the DUGKS has favorable stability properties in all transport regimes and especially low computational cost in the near-continuum stiff regime. In recent methodology developments, the DUGKS has been applied to model phonon transport problem based on the PBE under SMRT approximation [33-35]. Inspired by these works, we intend to develop the counterpart of DUGKS for the PBE under Callaway's model in this work. With the DUGKS scheme, it thus becomes possible to investigate the dynamics of second sound in graphene in a systematic way. Therefore, our work will promote both the development of solution methodology for PBE and the understanding of physics of the less understood transient phonon hydrodynamics.

The remainder of this paper is organized as follows. In Section 2, the PBE under Callaway's dual relaxation model as well as the DUGKS numerical scheme will be presented in details. In Section 3, two representative numerical cases of heat transport in graphene ribbon are tested to validate the present methodology. The DUGKS scheme is then applied to model two cases of second sound propagation as the paradigms of experimental detection in Section 4. The concluding remarks are finally outlined in Section 5.

## 2. Mathematical model and numerical methods

### 2.1 Phonon Boltzmann equation

The PBE describes the spatial-temporal evolution of the number density distribution function of an ensemble of phonons that is subject to advection and scattering. The PBE under Callaway's dual relaxation model has the form [27]:

$$\frac{\partial f(\mathbf{x},\mathbf{k},p,t)}{\partial t}+\mathbf{v}_{\mathbf{k}p}\cdot\nabla f(\mathbf{x},\mathbf{k},p,t)=\frac{f_{\mathrm{R}}^{\mathrm{eq}}(\mathbf{x},\mathbf{k},p,t)-f(\mathbf{x},\mathbf{k},p,t)}{\tau_{\mathrm{R}}}+\frac{f_{\mathrm{N}}^{\mathrm{eq}}(\mathbf{x},\mathbf{k},p,t)-f(\mathbf{x},\mathbf{k},p,t)}{\tau_{\mathrm{N}}} \tag{1}$$

where $f(\mathbf{x}, \mathbf{k}, p, t)$ is number density distribution of phonons around the position $\mathbf{x}$, the phonon wave-vector $\mathbf{k}$ for phonon polarization branch $p$ and time $t$. For simplification, the arguments ($\mathbf{x}$, $\mathbf{k}$, $p$, $t$) is omitted hereafter. $\mathbf{v}_{\mathbf{k},p}=\nabla_{\mathbf{k}}\omega_{\mathbf{k},p}$ is the phonon group velocity ($\omega_{\mathbf{k},p}$ is the phonon angular frequency). The subscript $\mathbf{k},p$ denotes a phonon mode in the $p$th phonon branch with a momentum $\hbar\mathbf{k}$ ($\hbar$ being the reduced Planck's constant) and indicates that group velocity is a function of wave-vector and polarization. $f_{\mathrm{R}}^{\mathrm{eq}}$ and $f_{\mathrm{N}}^{\mathrm{eq}}$ are the equilibrium Bose-Einstein distribution for resistive processes and displaced Bose-Einstein distribution for normal processes, respectively. The two local pseudo-equilibrium distribution functions are defined mathematically as:

$$f_{\mathrm{R}}^{\mathrm{eq}} = \frac{1}{\exp\left(\dfrac{\hbar\omega_{\mathbf{k},p}}{k_{\mathrm{B}}T_{\mathrm{R}}}\right)-1} \tag{2}$$

$$f_{\mathrm{N}}^{\mathrm{eq}} = \frac{1}{\exp\left(\dfrac{\hbar\omega_{\mathbf{k},p}-\hbar\mathbf{k}\cdot\mathbf{u}}{k_{\mathrm{B}}T_{\mathrm{N}}}\right)-1} \tag{3}$$

where $T_{\mathrm{R}}$ and $T_{\mathrm{N}}$ are the local pseudo-temperature for normal and resistive processes, respectively. $\mathbf{u}$ is a macroscopic phonon drift velocity. $k_{\mathrm{B}}$ is Boltzmann's constant. These macroscopic variables are determined by conservation principles as to be explained later. $\tau_{\mathrm{N}}$ is the phonon relaxation times due to normal scattering processes. $\tau_{\mathrm{R}}$ is the total intrinsic resistive relaxation time accounting for all phonon scattering processes that destroy the crystal momentum, and is given according to Matthiessen's rule as,

$$\frac{1}{\tau_{\mathrm{R}}} = \frac{1}{\tau_{\mathrm{U}}} + \frac{1}{\tau_{\mathrm{iso}}} \tag{4}$$

where $\tau_{\mathrm{U}}, \tau_{\mathrm{iso}}$ refers to the relaxation times for umklapp scattering and isotope scattering respectively. In general, the relaxation times $\tau_{\mathrm{N}}$ and $\tau_{\mathrm{U}}$ have strong dependences on both phonon frequency $\omega_{\mathbf{k},p}$ and the local temperature $T$. Phonon-isotope scattering rate is explicitly dependent on the phonon frequency.

For convenience, the phonon Boltzmann equation (1) is usually rewritten into a deviational energy form as below:

$$\frac{\partial g}{\partial t} + \mathbf{v}_{\mathbf{k},p}\cdot\nabla g = \frac{g_{\mathrm{R}}^{\mathrm{eq}}-g}{\tau_{\mathrm{R}}} + \frac{g_{\mathrm{N}}^{\mathrm{eq}}-g}{\tau_{\mathrm{N}}} \tag{5}$$

where the deviational energy distribution function is introduced as [36]:

$$g = \hbar\omega_{\mathbf{k},p}\left[f - f_{\mathrm{R}}^{\mathrm{eq}}\left(T_{\mathrm{ref}}\right)\right] \tag{6}$$

in which $T_{\mathrm{ref}}$ is the constant reference temperature throughout the system. Assuming the temperature difference of the system $\Delta T$ and the phonon drift velocity $\mathbf{u}$ satisfy the following conditions: $\mathbf{k}\cdot\mathbf{u}\leq\omega_{\mathbf{k},p}$, $\Delta T/T_{\mathrm{ref}}\ll 1$, the pseudo-equilibrium deviational energy distribution functions are linearized to:

$$g_{\mathrm{R}}^{\mathrm{eq}} = \hbar\omega_{\mathbf{k},p}\left[f_{\mathrm{R}}^{\mathrm{eq}} - f_{\mathrm{R}}^{\mathrm{eq}}\left(T_{\mathrm{ref}}\right)\right] \approx C_{\mathbf{k},p}\left(T_{\mathrm{R}} - T_{\mathrm{ref}}\right) \tag{7}$$

$$g_{\mathrm{N}}^{\mathrm{eq}}\left(T_{\mathrm{N}}\right)=\hbar\omega_{\mathbf{k},p}\left[f_{\mathrm{N}}^{\mathrm{eq}}-f_{\mathrm{R}}^{\mathrm{eq}}\left(T_{\mathrm{ref}}\right)\right]\approx C_{\mathbf{k},p}\left(T_{\mathrm{N}}-T_{\mathrm{ref}}+T_{\mathrm{N}}\frac{\mathbf{k}\cdot\mathbf{u}}{\omega}\right) \tag{8}$$

where the mode-dependent heat capacity is expressed as

$$C_{\mathbf{k},p}=\hbar\omega_{\mathbf{k},p}\left.\frac{\partial f_{\mathrm{R}}^{\mathrm{eq}}}{\partial T}\right|_{T=T_{\mathrm{ref}}}=f_{\mathrm{R}}^{\mathrm{eq}}\left(T_{\mathrm{ref}}\right)\left[f_{\mathrm{R}}^{\mathrm{eq}}\left(T_{\mathrm{ref}}\right)+1\right]\frac{\hbar\omega_{\mathbf{k},p}}{k_{\mathrm{B}}T_{\mathrm{ref}}^{2}}. \tag{9}$$

These linearized expressions are useful in the computation of macroscopic variables as to be discussed below. Small temperature difference also suggests that the phonon relaxation times can be approximately evaluated at the reference temperature $T_{\mathrm{ref}}$. In general, the two local pseudo-temperatures and the drift velocity are calculated based on the energy and momentum conservation of resistive and normal processes [27]:

$$\sum_p\int\frac{g_{\mathrm{R}}^{\mathrm{eq}}-g}{\tau_{\mathrm{R}}}\frac{\mathrm{d}\mathbf{k}}{D_0}=0 \tag{10}$$

$$\sum_p\int\frac{g_{\mathrm{N}}^{\mathrm{eq}}-g}{\tau_{\mathrm{N}}}\frac{\mathrm{d}\mathbf{k}}{D_0}=0 \tag{11}$$

$$\sum_p\int\frac{g_{\mathrm{N}}^{\mathrm{eq}}-g}{\tau_{\mathrm{N}}}\frac{\mathbf{k}}{\omega}\frac{\mathrm{d}\mathbf{k}}{D_0}=0 \tag{12}$$

where $D_0$ is a parameter dependent on the system dimension. For three-dimensional materials, $D_0=(2\pi)^3$ such that Eqs. (10), (11) and (12) are 3-D integrals over the wave-vector space. For single-layer two-dimensional materials, $D_0=(2\pi)^2h_0$ with the layer thickness $h_0$; and Eqs. (10), (11)and (12) are 2-D integrals. Substitution of Eq. (7) into Eq. (10) yields:

$$T_{\mathrm{R}}-T_{\mathrm{ref}}=\sum_p\int\frac{g}{\tau_{\mathrm{R}}}\frac{\mathrm{d}\mathbf{k}}{D_0}\Bigg/\sum_p\int\frac{C_{\mathbf{k},p}}{\tau_{\mathrm{R}}}\frac{\mathrm{d}\mathbf{k}}{D_0} \tag{13}$$

Again, substituting Eq. (8) into Eqs. (11) and (12), respectively, give rise to:

$$T_{\mathrm{N}}-T_{\mathrm{ref}}=\sum_p\int\frac{g}{\tau_{\mathrm{N}}}\frac{\mathrm{d}\mathbf{k}}{D_0}\Bigg/\sum_p\int\frac{C_{\mathbf{k},p}}{\tau_{\mathrm{N}}}\frac{\mathrm{d}\mathbf{k}}{D_0} \tag{14}$$

$$T_{\mathrm{N}}\mathbf{u}\cdot\sum_p\int\frac{C_{\mathbf{k},p}\mathbf{k}\mathbf{k}}{\tau_{\mathrm{N}}\omega^2}\frac{\mathrm{d}\mathbf{k}}{D_0}=\sum_p\int\frac{g}{\tau_{\mathrm{N}}}\frac{\mathbf{k}}{\omega}\frac{\mathrm{d}\mathbf{k}}{D_0} \tag{15}$$

The local temperature $T$ is defined as a measure of the local energy density for which the equilibrium energy density of phonons matches the non-equilibrium energy density. The linearization of equilibrium energy density gives the local deviational temperature as the ratio of the non-equilibrium energy density of phonons to the heat capacity, i.e.

$$T - T_{\text{ref}} = \sum_{p} \int g \frac{\mathbf{dk}}{D_0} \bigg/ \sum_{p} \int C_{\mathbf{k},p} \frac{\mathbf{dk}}{D_0} \tag{16}$$

The heat flux of interest is defined as follows,

$$\mathbf{q} = \sum_{p} \int g \mathbf{v}_{\mathbf{k},p} \frac{\mathbf{dk}}{D_0} \tag{17}$$

The characteristics of the phonon transport in crystalline materials depend on the relative strength of the normal processes and resistive processes. When the intrinsic resistive scattering (umklapp scattering and isotope scattering) are predominant, the system approaches the diffusive regime. When the normal scattering dominates over other types of scattering processes, phonon transport is hydrodynamic. In this hydrodynamic regime, phonons exhibit a collective motion with a nonzero drift velocity when they are subjected to a temperature gradient, analogous to a viscous fluid system driven by a pressure gradient. Due to small relaxation times comparing to the characteristic time, the PBE becomes stiff in the diffusive regime or the hydrodynamic regime and represents a computational challenge. Generally, the time discretization is a key point for a numerical model within stiff regime. For traditional explicit scheme, the time step is restricted by the usual parabolic Courant-Friedrichs-Lewy (CFL) condition [37,38]. The use of large time steps may cause numerical instability and the loss of numerical accuracy. Fully implicit time integration technique has no such restriction on the time step but has to invert a linear system, which is computationally expensive in high-dimensional cases. Moreover, fully implicit scheme usually induces more numerical diffusion. To tackle these issues, it is indispensable to develop special time discretization schemes, which will be discussed in the following section.

### 2.2 Discrete unified gas kinetic scheme for the PBE

Inspired by Refs. [33,34], we construct a robust and efficient grid-based numerical method for the PBE under Callaway's dual relaxation model based on the discrete unified kinetic scheme in this subsection. Firstly, we deal with the discretization of wave-vector space (or called the first Brillouin zone), real-space and time. The first Brillouin zone is usually discretized into a uniform grid of $\mathbf{k}$ points centered around the Γ point. Based on a certain quadrature rule, the integration over the first Brillouin zone is transformed into a discrete summation of all discrete $\mathbf{k}$ points. For an assumed isotropic wave-vector space, the phonon frequency only depends on the magnitude of the wave-vector $\mathbf{k}$ whereas independent on its direction. Thus the integration can be calculated by summing over all discrete frequencies and directions of wave-vector, as done in Ref. [22]. For each discrete **k** point, the corresponding discrete frequency and group velocity are calculated from the analytical isotropic phonon dispersion. The scattering rates can be obtained by first principle calculations or just using the empirical relation. Without loss of generality, we present the numerical scheme on the general wave-vector space in below and let $C_{m,p}$, $\mathbf{v}_{m,p}$, $\omega_{m,p}$ and $W_{m,p}$ denote the discrete mode heat capacity, discrete group velocity, discrete frequency and the corresponding quadrature weights at a given discrete phonon mode ($\mathbf{k}_m$, $p$). The real-space domain is discretized into a number of control cells by finite volume method. Furthermore, the time domain is decomposed into equidistant discrete times $t_1$, $t_2$,…,$t_n$ with time step $\Delta t=t_{n+1}-t_n$. For convenience to extend the DUGKS scheme for the PBE under SMRT approximation [32, 33] for the Callaway's model, we rewrite the discrete ordinate form of Eq.(5) as

$$\frac{\partial}{\partial t}g\left(\mathbf{x},\mathbf{k}_m,p,t\right)+\mathbf{v}_{m,p}\cdot\nabla g\left(\mathbf{x},\mathbf{k}_m,p,t\right)=\frac{g^{\mathrm{eq}}\left(\mathbf{x},\mathbf{k}_m,p,t\right)-g\left(\mathbf{x},\mathbf{k}_m,p,t\right)}{\tau_{\mathrm{C}}} \tag{18}$$

where $\tau_{\mathrm{C}}=\tau_{\mathrm{R}}\tau_{\mathrm{N}}/(\tau_{\mathrm{R}}+\tau_{\mathrm{N}})$ is the total relaxation time and an effective equilibrium distribution function is fully expressed as,

$$g^{\mathrm{eq}}=\frac{\tau_{\mathrm{C}}}{\tau_{\mathrm{R}}}g_{\mathrm{R}}^{\mathrm{eq}}+\frac{\tau_{\mathrm{C}}}{\tau_{\mathrm{N}}}g_{\mathrm{N}}^{\mathrm{eq}} \tag{19}$$

Here, the effective equilibrium distribution function is introduced just for the convenience of numerical solution of Eq. (5). It doesn't mean that the resistive process and normal process can be treated in an equal level as in the SMRT approximation. Thus Eq. (18) shares the similar mathematical form as that of the PBE under SMRT approximation. In this way, the same numerical strategy in Ref. [34] can be borrowed here in a straightforward way. As a finite-volume-based method, the calculations of the cell-averaged distribution function and the numerical flux at cell interface are two key ingredients in the DUGKS scheme, as to be illustrated in the following subsection 2.2.1 and 2.2.2, respectively.

### 2.2.1 The evolution of the cell-averaged distribution function

Analogously as in Ref. [34], we integrate Eq. (18) over one control cell with volume $V_i$ and integrating from time $t_n$ to $t_{n+1}$ with the midpoint rule for the advection term and the trapezoidal rule for collision term. Thus we get an update equation for the averaged distribution function in cell $i$,

$$\begin{aligned}g\left(\mathbf{x}_i,\mathbf{k}_m,p,t_{n+1}\right)=g\left(\mathbf{x}_i,\mathbf{k}_m,p,t_n\right)+\frac{\Delta t}{2\tau_{\mathrm{C}}}\left[g^{\mathrm{eq}}\left(\mathbf{x}_i,\mathbf{k}_m,p,t_{n+1}\right)-g\left(\mathbf{x}_i,\mathbf{k}_m,p,t_{n+1}\right)\right]\\+\frac{\Delta t}{2\tau_{\mathrm{C}}}\left[g^{\mathrm{eq}}\left(\mathbf{x}_i,\mathbf{k}_m,p,t_n\right)-g\left(\mathbf{x}_i,\mathbf{k}_m,p,t_n\right)\right]-\frac{\Delta t}{V_i}F\left(\mathbf{x}_i,\mathbf{k}_m,p,t_{n+1/2}\right)\end{aligned} \tag{20}$$

where $g(\mathbf{x}_i,\mathbf{k}_\mathrm{m},p,t_n)$ denotes the averaged distribution function in cell $i$, around the discrete phonon mode ($\mathbf{k}_m$, $p$), and at time step $t_n$. $F(\mathbf{x}_i,\mathbf{k}_\mathrm{m},p,t_{n+1/2})$ are the total numerical flux across all control surfaces of cell $i$ at $t_{n+1/2}=t_n+\Delta t/2$. This time integration scheme makes the present scheme numerically stable and ensures second-order accuracy in time discretization. Since the effective equilibrium distribution function $g^{\mathrm{eq}}$ depends on the macroscopic variables ($T_\mathrm{R}$, $T_\mathrm{N}$, $\mathbf{u}$), the distribution function at the next time step cannot be obtained directly. To avoid the implicit computation of Eq.(20), we introduce two intermediate distribution functions,

$$\tilde{g}=g-\frac{\Delta t}{2\tau_\mathrm{C}}\left(g^{\mathrm{eq}}-g\right)=g-\frac{\Delta t}{2}\left(\frac{g_\mathrm{R}^{\mathrm{eq}}-g}{\tau_\mathrm{R}}+\frac{g_\mathrm{N}^{\mathrm{eq}}-g}{\tau_\mathrm{N}}\right) \tag{21}$$

$$\tilde{g}^{+}=g+\frac{\Delta t}{2\tau_\mathrm{C}}\left(g^{\mathrm{eq}}-g\right)=g+\frac{\Delta t}{2}\left(\frac{g_\mathrm{R}^{\mathrm{eq}}-g}{\tau_\mathrm{R}}+\frac{g_\mathrm{N}^{\mathrm{eq}}-g}{\tau_\mathrm{N}}\right) \tag{22}$$

As a consequence, Eq. (20) can be rewritten as

$$\tilde{g}\left(\mathbf{x}_i,\mathbf{k}_m,p,t_{n+1}\right)=\tilde{g}^{+}\left(\mathbf{x}_i,\mathbf{k}_m,p,t_n\right)-\frac{\Delta t}{V_i}F\left(\mathbf{x}_i,\mathbf{k}_m,p,t_{n+1/2}\right) \tag{23}$$

In practical computation, we evolve the intermediate distribution function $\tilde{g}$ instead of the original distribution function $g$.

In order to compute the time-dependent macroscopic variables, we reformulate Eq.(21) as

$$g=\frac{2\tau_\mathrm{C}}{2\tau_\mathrm{C}+\Delta t}\tilde{g}+\frac{\Delta t}{2\tau_\mathrm{C}+\Delta t}\frac{\tau_\mathrm{C}}{\tau_\mathrm{R}}g_\mathrm{R}^{\mathrm{eq}}+\frac{\Delta t}{2\tau_\mathrm{C}+\Delta t}\frac{\tau_\mathrm{C}}{\tau_\mathrm{N}}g_\mathrm{N}^{\mathrm{eq}} \tag{24}$$

In the present work, we assume the first Brillouin zone is highly symmetrical such that the integration over wave-vector space vanishes if the integrand is an odd function of $\mathbf{k}$. With this quadrature property, substitutions of Eq. (24) into Eqs. (10) and (11), respectively give rise to a linear equation set with respect to $T_\mathrm{R}$ and $T_\mathrm{N}$

$$\begin{cases} A_{11}\left[T_\mathrm{R}\left(\mathbf{x}_i,t_n\right)-T_{\mathrm{ref}}\right]+A_{12}\left[T_\mathrm{N}\left(\mathbf{x}_i,t_n\right)-T_{\mathrm{ref}}\right]=\sum_p\sum_m\frac{2\tau_\mathrm{C}}{\left(2\tau_\mathrm{C}+\Delta t\right)\tau_\mathrm{R}}\frac{W_{m,p}}{D_0}\tilde{g}\left(\mathbf{x}_i,\mathbf{k}_m,p,t_n\right) \\ A_{21}\left[T_\mathrm{R}\left(\mathbf{x}_i,t_n\right)-T_{\mathrm{ref}}\right]+A_{22}\left[T_\mathrm{N}\left(\mathbf{x}_i,t_n\right)-T_{\mathrm{ref}}\right]=\sum_p\sum_m\frac{2\tau_\mathrm{C}}{\left(2\tau_\mathrm{C}+\Delta t\right)\tau_\mathrm{N}}\frac{W_{m,p}}{D_0}\tilde{g}\left(\mathbf{x}_i,\mathbf{k}_m,p,t_n\right) \end{cases} \tag{25}$$

where the coefficients are fully expressed as

$$\begin{aligned} &A_{11}=\sum_p\sum_m\left(1-\frac{\Delta t}{2\tau_\mathrm{C}+\Delta t}\frac{\tau_\mathrm{C}}{\tau_\mathrm{R}}\right)\frac{C_{m,p}}{\tau_\mathrm{R}}\frac{W_{m,p}}{D_0},A_{22}=\sum_p\sum_m\left(1-\frac{\Delta t}{2\tau_\mathrm{C}+\Delta t}\frac{\tau_\mathrm{C}}{\tau_\mathrm{N}}\right)\frac{C_{m,p}}{\tau_\mathrm{N}}\frac{W_{m,p}}{D_0}, \\ &A_{12}=A_{21}=-\sum_p\sum_m\frac{C_{m,p}\Delta t}{\left(2\tau_\mathrm{C}+\Delta t\right)\left(\tau_\mathrm{R}+\tau_\mathrm{N}\right)}\frac{W_{m,p}}{D_0}. \end{aligned} \tag{26}$$

Note that the coefficients $A_{11}$, $A_{12}$, $A_{21}$, and $A_{22}$ don't depend on the distribution function and macroscopic variables, and satisfy $A_{11}\cdot A_{22}-A_{12}\cdot A_{21}\neq 0$. As the distribution function $\tilde{g}\left(\mathbf{x}_i,\mathbf{k}_m,p,t_n\right)$ is known before each update, it is easy to obtain $T_{\mathrm{R}}$ and $T_{\mathrm{N}}$ at the present time step through solving the linear equation set Eq. (25) by matrix inversion. Similarly, by substituting Eq. (24) into Eq.(12), we obtain that

$$T_{\mathrm{N}}\left(\mathbf{x}_i,t_n\right)\mathbf{u}\left(\mathbf{x}_i,t_n\right)\cdot\sum_p\sum_m\frac{C_{m,p}\mathbf{k}_m\mathbf{k}_m}{\tau_{\mathrm{N}}\omega_{m,p}^2}\left(1-\frac{\Delta t}{2\tau_{\mathrm{C}}+\Delta t}\frac{\tau_{\mathrm{C}}}{\tau_{\mathrm{N}}}\right)\frac{W_{m,p}}{D_0}=\sum_p\sum_m\frac{2\tau_{\mathrm{C}}\tilde{g}\left(\mathbf{x}_i,\mathbf{k}_m,p,t_n\right)}{\left(2\tau_{\mathrm{C}}+\Delta t\right)\tau_{\mathrm{N}}}\frac{\mathbf{k}_m}{\omega_{m,p}}\frac{W_{m,p}}{D_0}, \tag{27}$$

from which the macroscopic drift velocity $\mathbf{u}$ at the present time step is calculated. Once the local pseudo-temperature $T_{\mathrm{R}}$ and $T_{\mathrm{N}}$, and the drift velocity $\mathbf{u}$ are obtained, the local pseudo-equilibrium distribution functions in Eq. (24) can be determined based on Eqs. (7) and (8). The original distribution function is then determined by Eq. (24) and the local temperature and heat flux at the present time step are calculated based on Eqs. (16) and (17), respectively.

2.2.2 The evaluation of numerical flux at cell interface

The numerical flux across the interface can be obtained by using the Gauss theory as

$$F\left(\mathbf{x}_i,\mathbf{k}_m,p,t_{n+1/2}\right)=\int_{V_i}\mathbf{v}_{m,p}\cdot\nabla g\left(\mathbf{x},\mathbf{k}_m,p,t_{n+1/2}\right)\mathrm{d}V=\sum_{j\in N(i)}\left(\mathbf{v}_{m,p}\cdot\mathbf{n}_{ij}\right)S_{ij}g\left(\mathbf{x}_{ij},\mathbf{k}_m,p,t_{n+1/2}\right) \tag{28}$$

where $N(i)$ denotes the set of facing neighbor cells of cell $i$; $S_{ij}$ is the area of the cell interface shared by cell $i$ and cell $j$; and $\mathbf{n}_{ij}$ is the unit normal vector of the cell interface directing from cell $i$ to cell $j$. In Eq. (28), the non-equilibrium distribution function $g(\mathbf{x}_{ij},\mathbf{k}_m,p,t_{n+1/2})$ at the cell interface should be evaluated properly to accurately calculate the numerical flux. In traditional discrete ordinate method and finite volume method, upwind schemes are usually employed to construct the interface distribution function for their excellent numerical stability [22]. However, these upwind techniques only consider the streaming process of phonon transport at the cell interface and may induce extra numerical dissipation. In the DUGKS scheme, the interface distribution function is solved by the method of characteristics, where both the streaming process and scattering process are taken into account. To be specific, we integrate Eq. (18) along the characteristic line within a half time step with the end point located at the cell interface $\mathbf{x}_{ij}$,

$$\begin{aligned}g\left(\mathbf{x}_{ij},\mathbf{k}_m,p,t_{n+1/2}\right)=g\left(\mathbf{x}_{ij}-\mathbf{v}_{m,p}\Delta t/2,\mathbf{k}_m,p,t_n\right)+\frac{\Delta t}{4\tau_{\mathrm{C}}}\left[g^{\mathrm{eq}}\left(\mathbf{x}_{ij},\mathbf{k}_m,p,t_{n+1/2}\right)-g\left(\mathbf{x}_{ij},\mathbf{k}_m,p,t_{n+1/2}\right)\right]\\+\frac{\Delta t}{4\tau_{\mathrm{C}}}\left[g^{\mathrm{eq}}\left(\mathbf{x}_{ij}-\mathbf{v}_{m,p}\Delta t/2,\mathbf{k}_m,p,t_n\right)-g\left(\mathbf{x}_{ij}-\mathbf{v}_{m,p}\Delta t/2,\mathbf{k}_m,p,t_n\right)\right]\end{aligned} \tag{29}$$

where the time integration of the term in the right side of Eq. (18) is approximated by the trapezoidal rule. Similar to the treatment in Eq. (20), Eq. (29) can be rewritten in an explicit form by introducing another two intermediate distribution functions,

$$\bar{g}\left(\mathbf{x}_{ij},\mathbf{k}_m,p,t_{n+1/2}\right)=\bar{g}^{+}\left(\mathbf{x}_{ij}-\mathbf{v}_{m,p}\Delta t/2,\mathbf{k}_m,p,t_n\right) \tag{30}$$

where

$$\bar{g}=g-\frac{\Delta t}{4\tau_{\mathrm{C}}}\left(g^{\mathrm{eq}}-g\right)=g-\frac{\Delta t}{4}\left(\frac{g_{\mathrm{R}}^{\mathrm{eq}}-g}{\tau_{\mathrm{R}}}+\frac{g_{\mathrm{N}}^{\mathrm{eq}}-g}{\tau_{\mathrm{N}}}\right) \tag{31}$$

$$\bar{g}^{+}=g+\frac{\Delta t}{4\tau_{\mathrm{C}}}\left(g^{\mathrm{eq}}-g\right)=g+\frac{\Delta t}{4}\left(\frac{g_{\mathrm{R}}^{\mathrm{eq}}-g}{\tau_{\mathrm{R}}}+\frac{g_{\mathrm{N}}^{\mathrm{eq}}-g}{\tau_{\mathrm{N}}}\right). \tag{32}$$

The spatial point $\mathbf{x}_{ij}$-$\mathbf{v}_{m,p}\Delta t/2$ in Eq. (30) may not be located in the cell center, and $\bar{g}^{+}\left(\mathbf{x}_{ij}-\mathbf{v}_{m,p}\Delta t/2,\mathbf{k}_m,p,t_n\right)$ is approximated by interpolation as

$$\begin{aligned}&\bar{g}^{+}\left(\mathbf{x}_{ij}-\mathbf{v}_{m,p}\Delta t/2,\mathbf{k}_m,p,t_n\right)\\&=\bar{g}^{+}\left(\mathbf{x}_J,\mathbf{k}_m,p,t_n\right)+\left(\mathbf{x}_{ij}-\mathbf{x}_J-\mathbf{v}_{m,p}\Delta t/2\right)\cdot\sigma\left(\mathbf{x}_J,\mathbf{k}_m,p,t_n\right),J=\begin{cases}i & \mathbf{v}_{m,p}\cdot\mathbf{n}_{ij}>0\\ j & \text{otherwise}\end{cases}\end{aligned} \tag{33}$$

where $\sigma$ is the discrete gradient of the intermediate distribution function $\bar{g}^{+}$, which can be constructed smoothly or using flux limiters in the practical simulations. More details about the gradient construction are available in Refs. [32,33].

The calculation of the macroscopic variables from $\bar{g}\left(\mathbf{x},\mathbf{k},p,t\right)$ is similar to that from $\tilde{g}\left(\mathbf{x},\mathbf{k},p,t\right)$. $T_{\mathrm{R}}$ and $T_{\mathrm{N}}$ are obtained by solving the following linear equation set,

$$\begin{cases}B_{11}\left[T_{\mathrm{R}}\left(\mathbf{x}_{ij},t_{n+1/2}\right)-T_{\mathrm{ref}}\right]+B_{12}\left[T_{\mathrm{N}}\left(\mathbf{x}_{ij},t_{n+1/2}\right)-T_{\mathrm{ref}}\right]=\sum_p\sum_m\frac{4\tau_{\mathrm{C}}}{\left(4\tau_{\mathrm{C}}+\Delta t\right)\tau_{\mathrm{R}}}\frac{W_{m,p}}{D_0}\bar{g}\left(\mathbf{x}_{ij},\mathbf{k}_m,p,t_{n+1/2}\right)\\ B_{21}\left[T_{\mathrm{R}}\left(\mathbf{x}_{ij},t_{n+1/2}\right)-T_{\mathrm{ref}}\right]+B_{22}\left[T_{\mathrm{N}}\left(\mathbf{x}_{ij},t_{n+1/2}\right)-T_{\mathrm{ref}}\right]=\sum_p\sum_m\frac{4\tau_{\mathrm{C}}}{\left(4\tau_{\mathrm{C}}+\Delta t\right)\tau_{\mathrm{N}}}\frac{W_{m,p}}{D_0}\bar{g}\left(\mathbf{x}_{ij},\mathbf{k}_m,p,t_{n+1/2}\right)\end{cases} \tag{34}$$

where the coefficients are given by

$$\begin{aligned}&B_{11}=\sum_p\sum_m\left(1-\frac{\Delta t}{4\tau_{\mathrm{C}}+\Delta t}\frac{\tau_{\mathrm{C}}}{\tau_{\mathrm{R}}}\right)\frac{C_{m,p}}{\tau_{\mathrm{R}}}\frac{W_{m,p}}{D_0},B_{22}=\sum_p\sum_m\left(1-\frac{\Delta t}{4\tau_{\mathrm{C}}+\Delta t}\frac{\tau_{\mathrm{C}}}{\tau_{\mathrm{N}}}\right)\frac{C_{m,p}}{\tau_{\mathrm{N}}}\frac{W_{m,p}}{D_0},\\&B_{12}=B_{21}=-\sum_p\sum_m\frac{C_{m,p}\Delta t}{\left(4\tau_{\mathrm{C}}+\Delta t\right)\left(\tau_{\mathrm{R}}+\tau_{\mathrm{N}}\right)}\frac{W_{m,p}}{D_0}.\end{aligned} \tag{35}$$

The phonon drift velocity $\mathbf{u}$ is obtained by solving the following equation,

$$T_{\mathrm{N}}\left(\mathbf{x}_{ij},t_{n+1/2}\right)\mathbf{u}\left(\mathbf{x}_{ij},t_{n+1/2}\right)\cdot\sum_p\sum_m\frac{C_{m,p}\mathbf{k}_m\mathbf{k}_m}{\tau_{\mathrm{N}}\omega_{m,p}^2}\left(1-\frac{\Delta t}{4\tau_{\mathrm{C}}+\Delta t}\frac{\tau_{\mathrm{C}}}{\tau_{\mathrm{N}}}\right)\frac{W_{m,p}}{D_0}=\sum_p\sum_m\frac{4\tau_{\mathrm{C}}\bar{g}\left(\mathbf{x}_{ij},\mathbf{k}_m,p,t_{n+1/2}\right)}{\left(4\tau_{\mathrm{C}}+\Delta t\right)\tau_{\mathrm{N}}}\frac{\mathbf{k}_m}{\omega_{m,p}}\frac{W_{m,p}}{D_0} \tag{36}$$

Once the pseudo-temperatures and drift velocity are obtained, the pseudo-equilibrium distribution functions at a cell interface are fully determined. Thus the original distribution function can be obtained by

$$g = \frac{4\tau_{\mathrm{C}}}{4\tau_{\mathrm{C}} + \Delta t}\bar{g} + \frac{\Delta t}{4\tau_{\mathrm{C}} + \Delta t}g^{\mathrm{eq}} \tag{37}$$

In order to avoid implicit iteration in the calculation of cell-averaged distribution function and interface distribution function, four intermediate distribution functions have been introduced. To complete the whole iteration procedure, the relations between these intermediate distribution functions and the determination of time step are explained as follow. Firstly, we focus the intermediate distribution function $\bar{g}^{+}$ and $\tilde{g}^{+}$. It is easy to get the relation between $\bar{g}^{+}$ and $\tilde{g}$ from Eqs. (21) and (32) through a mathematical transformation,

$$\bar{g}^{+} = \frac{4\tau_{\mathrm{C}} - \Delta t}{4\tau_{\mathrm{C}} + 2\Delta t}\tilde{g} + \frac{3\Delta t}{4\tau_{\mathrm{C}} + 2\Delta t}g^{\mathrm{eq}} \tag{38}$$

By combining Eqs. (21), (22) and (32), we can also obtain that

$$\tilde{g}^{+} = \frac{4}{3}\bar{g}^{+} - \frac{1}{3}\tilde{g}. \tag{39}$$

In order to avoid extra-interpolation in Eq. (33), the maximum distance of phonon free transport within half a time step should be smaller than the half of the cell size, which determines the time step in DUGKS simulation. This restriction can be also rewritten as the form of the CFL condition,

$$\Delta t = \alpha \frac{\Delta x_{\min}}{v_{\mathrm{g}}^{\max}} \tag{40}$$

where $0<\alpha<1$ is the CFL number, $\Delta x_{\min}$ is the minimal mesh size, $v_{\mathrm{g}}^{\max}$ is the maximum value of phonon group velocity. Eq. (40) infers that the time step in the present method is not limited by the phonon relaxation time as in explicit temporal integration scheme. This allows the use of time and spatial steps that are independent of phonon relaxation time and MFP. This special time discretization scheme in the present method guarantees good numerical stability. In addition, all the computation is explicit and can be implemented in a local stencil instead of the whole grid system, resulting in low computational cost. The inclusion of the scattering process in numerical flux reconstruction endows the present method with higher robustness in physics. As a consequence, the present numerical method for the Callaway's dual relaxation model is asymptotic preserving. In other words, the present scheme is reduced to a consistent numerical approximation of the relevant macroscopic model (Fourier diffusion equation and phonon hydrodynamic equation) in different asymptotic regimes (diffusive limit and hydrodynamic limit) at the discrete level.

### 2.3 Boundary treatment

In the practical modeling of hydrodynamic phonon transport in a finite domain, boundary conditions are necessary for the determination of the deviational energy distribution function at the boundary. In general, three types of boundary condition are often encountered: (1) isothermal boundary condition (2) heat flux boundary condition or adiabatic boundary condition and (3) periodic heat flux boundary condition or periodic boundary condition.

An isothermal boundary is assumed to be analogous to an isothermal black surface for thermal radiation. Assuming fully absorbing condition, the deviation energy distribution function of phonons leaving such boundary located at $\mathbf{x}_\mathrm{B}$ is written as,

$$g\left(\mathbf{x}_\mathrm{B}\right)=C_{\mathbf{k},p}\left(T_\mathrm{B}-T_\mathrm{ref}\right), \mathbf{v}_{\mathbf{k},p}\cdot\mathbf{n}_\mathrm{B}>0 \tag{41}$$

where $T_\mathrm{B}$ and $\mathbf{n}_\mathrm{B}$ are the boundary temperature and the unit normal vector of the boundary pointing to the computational domain, respectively.

Heat flux boundary condition is a subject of considerable interest in many applications. To ensure no loss of energy at such boundaries, the phonons striking the boundary are reflected diffusely back into the computational domain without change of polarization and frequency. Additionally, extra energy is instantaneously added to these reflected phonons to exert a constant heat flux. Assuming $q_0$ denotes the net normal heat flux imposed on the boundary, the unknown distribution function for phonons reentering the computational domain is expressed mathematically as,

$$g\left(\mathbf{x}_\mathrm{B}\right)=\frac{q_0}{\sum_p\int_{\mathbf{v}_{\mathbf{k},p}\cdot\mathbf{n}_\mathrm{B}>0}\left(\mathbf{v}_{\mathbf{k},p}\cdot\mathbf{n}_\mathrm{B}\right)\frac{\mathrm{d}\mathbf{k}}{D_0}}-\frac{\int_{\mathbf{v}_{\mathbf{k},p}\cdot\mathbf{n}_\mathrm{B}<0}\left(\mathbf{v}_{\mathbf{k},p}\cdot\mathbf{n}_\mathrm{B}\right)\delta\left(\omega_{\mathbf{k},p}\right)g\left(\mathbf{x}_\mathrm{B}\right)\frac{\mathrm{d}\mathbf{k}}{D_0}}{\int_{\mathbf{v}_{\mathbf{k},p}\cdot\mathbf{n}_\mathrm{B}>0}\left(\mathbf{v}_{\mathbf{k},p}\cdot\mathbf{n}_\mathrm{B}\right)\delta\left(\omega_{\mathbf{k},p}\right)\frac{\mathrm{d}\mathbf{k}}{D_0}}, \mathbf{v}_{\mathbf{k},p}\cdot\mathbf{n}_\mathrm{B}>0 \tag{42}$$

where $\delta$ is the Dirac function. The adiabatic boundary condition can be treated as a special heat flux boundary condition with zero heat flux. Partially diffuse and partially specular scheme shall be adopted for more realistic boundary. A fully diffuse scheme is assumed here for simplicity and clarity of discussion. Such an assumption is usually valid for rough boundary and an elevated temperature scope where the phonon wavelength is much smaller than the characteristic size of boundary roughness.

The periodic heat flux boundary condition is designed to reduce the computational cost of heat transport in periodic nanostructures through an implementation of a constant temperature gradient within a periodic unit. Here the direction of the temperature gradient is given as from left side of to the right side of the computational domain. In the simulation of in-plane heat transport through an infinitely long sample, we artificially add two virtual periodic heat flux boundaries to reduce the infinite computation domain to a finite one. The deviational energy distribution function on the left boundary (LB) and the right boundary (RB) are respectively given by the following expressions,

$$g\left(\mathbf{x}_{\mathrm{LB}}\right)=g\left(\mathbf{x}_{\mathrm{RB}}\right)+C_{\mathbf{k},p}\left(T_{\mathrm{LB}}-T_{\mathrm{RB}}\right), \mathbf{v}_{\mathbf{k},p}\cdot\mathbf{n}_{\mathrm{LB}}>0 \tag{43}$$

$$g\left(\mathbf{x}_{\mathrm{RB}}\right)=g\left(\mathbf{x}_{\mathrm{LB}}\right)+C_{\mathbf{k},p}\left(T_{\mathrm{RB}}-T_{\mathrm{LB}}\right), \mathbf{v}_{\mathbf{k},p}\cdot\mathbf{n}_{\mathrm{RB}}>0 \tag{44}$$

where $T_{\mathrm{LB}}$ and $T_{\mathrm{RB}}$ are virtual temperatures at the boundaries and related to the given temperature gradient. Particularly, the Eqs. (43) and (44) are reduced to the periodic boundary condition when $T_{\mathrm{LB}}= T_{\mathrm{RB}}$.

### 2.4 **Computational sequence**

The computational procedure of DUGKS scheme for the PBE under Callaway's dual relaxation model is depicted in Fig.1 and the critical steps from $t_{\mathrm{n}}$ to $t_{\mathrm{n+1}}$ are listed as follows:

(a) Calculate the intermediate distribution function $\bar{g}^{+}\left(\mathbf{x}_i,\mathbf{k}_m,p,t_n\right)$ according to Eq. (38).

(b) Reconstruct the gradient of $\bar{g}^{+}\left(\mathbf{x}_i,\mathbf{k}_m,p,t_n\right)$ and calculate the interface distribution function $\bar{g}\left(\mathbf{x}_{ij},\mathbf{k}_m,p,t_{n+1/2}\right)$ by using Eqs. (30) and (33).

(c) Determine the macroscopic variables($T_{\mathrm{R}}^{n+1/2}$, $T_{\mathrm{N}}^{n+1/2}$, $\mathbf{u}^{n+1/2}$) at the interface by solving Eqs. (34) and (36), respectively, and then calculate the original distribution function $g\left(\mathbf{x}_{ij},\mathbf{k}_m,p,t_{n+1/2}\right)$ according to Eq. (37).

(d) Calculate the intermediate distribution function $\tilde{g}^{+}\left(\mathbf{x}_i,\mathbf{k}_m,p,t_n\right)$ according to Eq. (39) and update the cell-averaged distribution function $\tilde{g}\left(\mathbf{x}_i,\mathbf{k}_m,p,t_{n+1}\right)$ according to Eqs. (23) and (28).

(e) Calculate the macroscopic variables ($T_{\mathrm{R}}^{n+1}$, $T_{\mathrm{N}}^{n+1}$, $\mathbf{u}^{n+1}$, $T^{n+1}$, $\mathbf{q}^{n+1}$) at the next time step.

## 3. Methodology validation

To validate the DUGKS scheme developed in Section 2, we simulate two test cases of hydrodynamic phonon transport in two-dimensional material graphene, where the normal scattering plays an important role. The first Brillouin zone of graphene is very isotropic and assumed to be a 2-D circular region. Thus the integration over the wave-vector space is simplified into a double integration with respective to the magnitude of wave-vector $k$=|$\mathbf{k}$|∈[0, $k_{\max}$] ($k_{\max}$ = 1.5×10$^{10}$m$^{-1}$) and the angular variable $\theta\in$ [0,2$\pi$] (related to the wave-vector by $\mathbf{k}$=[$k\cos\theta$, $k\sin\theta$]). In this way, the phonon mode dependence is expressed in terms of frequency instead of wave-vector, i.e.

$$\frac{1}{(2\pi)^2 h_0}\sum_p \int G(\mathbf{x},\mathbf{k},p,t)\,\mathrm{d}\mathbf{k} = \frac{1}{2\pi h_0}\sum_p \int_0^{k_{\max}} \int_{2\pi} G(\mathbf{x},k,\theta,p,t)\,k\,\mathrm{d}\theta\,\mathrm{d}k$$
$$= \frac{1}{h_0}\sum_p \int_0^{\omega_{\max,p}} \int_{2\pi} G(\mathbf{x},\omega,\theta,p,t)\,D(\omega,p)\,\mathrm{d}\theta\,\mathrm{d}\omega \tag{45}$$

with the integrand $G(\mathbf{x}, \mathbf{k}, p, t)$ an arbitrary function dependent on the phonon mode, the phonon density of state $D(\omega, p)=k/(2\pi v_{\mathbf{k},p})$ and the maximum frequency $\omega_{\max,p}$ for phonon polarization $p$. Similar to the treatment in Ref. [22], the Gauss-Legendre (G-L) quadrature is adopted to discretize the frequency $\omega$ and the angular variable $\theta$, which allows to increase discrete directions and reduce the ray effect as much as possible. The frequency spectrum of longitudinal acoustic (LA), transverse acoustic (TA) and flexural acoustic (ZA) branches are discretized into $N_{\mathrm{LA}}$, $N_{\mathrm{TA}}$ and $N_{\mathrm{ZA}}$ nodes, respectively. The angular variable $\theta$ is discretized into $N_\theta$ nodes. With the discrete wave-vector, other discrete parameters (i.e. frequency, group velocity, relaxation time, etc.) can be obtained from the phonon dispersion relation and empirical formulas of scattering rates, which are available in Refs. [22,39] (see Appendix A). With the given phonon properties, the bulk thermal conductivity of graphene with natural abundance is evaluated to be $1.2844\times10^5$ W/(m·K) and $3.7605\times10^3$ W/(m·K) respectively at two relevant temperatures 50K and 300K in our later study (for an intuitive comparison, the bulk thermal conductivity of the usual semiconductor material silicon is 2600 W/(m·K) and 156 W/(m·K) respectively at 50K and 300K [40]). The high thermal conductivity of graphene due to the hydrodynamic transport mechanism makes it a good candidate for heat dissipation applications.

### 3.1 Transient heat conduction in an infinitely wide graphene ribbon

The first test case is 1D transient thermal transport in an infinitely wide graphene ribbon with natural abundance of $^{13}$C $c$=1.1%. The length of the ribbon is $L$. Initially, the temperature of the ribbon is kept uniformly at $T_{\mathrm{L}}$=299.5 K. At $t$=0, a sudden temperature rise is imposed at the left boundary such that $T_{\mathrm{H}}$ = 300.5 K. This heat transport process is solved by the present discrete unified gas kinetic scheme with a uniform grid $N_x$=100, spectral nodes $N_{\mathrm{LA}}=N_{\mathrm{TA}}=N_{\mathrm{ZA}}$=20, and angular resolution $N_\theta$=128. The time step is determined by the CFL condition with CFL number $\alpha$ set as 0.852 in the present simulation.

Fig. 2(a) and (b) display the results of the time-dependent evolution of temperature field and heat flux distribution, respectively. Since the analytical solution to this case is not available, the numerical result obtained by an explicit second-order upwind scheme (see Appendix B) [33] with mesh resolution $N_x$=1000 serves as a benchmark solution. A dense mesh utilized in the explicit second-order upwind scheme ensures grid-independent result. As an explicit method, the time step is determined by $\Delta t = \min(\Delta x_{\min}/v_{\mathrm{g}}^{\max}, \tau_{\mathrm{C}}^{\min})$.

However, the tie-up between the time step and the phonon relaxation time is released in the present DUGKS, which allows the use of large time step in the stiff regime with small relaxation time. For comparison, the time step for the explicit second-order upwind scheme is also written as $\Delta t = \beta\Delta x_{\min}/v_{g}^{\max}$ with a length-dependent parameter $\beta$. According to our numerical tests, the calculation of explicit second-order upwind scheme blows up as $\beta$>0.5 at $L$=0.1 μm. With the increase of ribbon length, $\beta$ should be smaller. In order to obtain stable solutions, the value of $\beta$ in the explicit second-order upwind scheme is set to be 0.213, 0.0213, 0.00213 for $L$=0.1 μm, 1 μm, 10 μm, respectively. For all length considered here ($L$=0.1 μm, 1 μm, 10 μm), the numerical results of DUGKS are closely consistent with the reference solutions, as shown in Fig. 2(a) and (b). The present DUGKS combines the advantage of lower costs of explicit schemes and favorable stability properties of implicit scheme, which ensures the overall computational accuracy and efficiency.

### 3.2 Steady-state heat conduction in an infinitely long graphene ribbon

To validate the present scheme for steady-state case, we consider two-dimensional heat transport in an infinitely long monolayer graphene ribbon with a width $W$ as shown in Fig. 3(a). A uniform temperature gradient is implemented along the $x$-direction (the longitudinal direction of the ribbon) by the periodic heat flux boundary condition introduced in Section 2.3. The transverse boundaries are assumed to be adiabatic and implemented by the diffuse scheme. There are generally two different strategies to obtain the steady-state solution. One is to directly tackle the steady-state transport equation, as in Ref. [22]. The other treats the steady-state case as the long-time limit of a transient case through advancing the transient solution of the transport equation step-by-step until the solution is time-independent [33]. The second strategy is carried out in the present DUGKS simulation of the steady-state case. The convergence criterion is set by the condition that the maximum relative difference of both local temperature and heat flux between two adjacent iteration steps decreases to be less than $10^{-8}$.

Thermal transport in graphene ribbon with natural abundance around room temperature ($T_{\text{ref}}$=300 K) is firstly simulated by the present DUGKS, with the DOM solution in Ref. [22] as a benchmark. After independence test, the same angular resolution $N_{\theta}$=96 and spectral nodes $N_{\text{LA}}=N_{\text{TA}}=N_{\text{ZA}}$=20 are used in all the simulations. In the DUGKS simulation, a uniform spatial mesh of $N_x \times N_y$=5×100 is used and the CFL number is fixed at 0.5 unless stated otherwise. A few spatial grids are used along the $x$-direction attributed to the periodic heat flux boundary. In the DOM simulation, a coarse and dense spatial mesh of $N_x \times N_y$=5×100 and $N_x \times N_y$=5×1000 are used separately. The latter dense mesh is expected to provide grid-independent solution as a benchmark for comparison. Fig. 3(b) displays the $x$-directional heat flux as a function of $y$ coordinate at different ribbon widths, where the heat flux is normalized by the bulk value in an infinitely wide graphene ribbon under the same uniform temperature gradient. The numerical results by the present DUGKS are in excellent agreement with the DOM solutions for all the ribbon widths. As a result, the width-dependent effective thermal conductivity normalized by the bulk value by the present DUGKS is also consistent with the DOM solution, as shown in Fig. 3(c). For comparison, the effective thermal conductivity obtained from the MC solution of PBE with the *ab initio* full scattering term [41] has also been included in Fig. 3(c). The overall good agreement indicates that the Callaway's dual relaxation model indeed provides a good approximation to the full scattering term. As expected, the effective thermal conductivity increases linearly with the width $W$ in the ballistic regime and then increases sub-linearly with $W$ in the quasi-ballistic regime until it approaches the bulk limit (at about $W$=100 μm).

In order to demonstrate the numerical performance of the present DUGKS in the hydrodynamic transport regime, we then consider the thermal transport in an isotopically pure graphene ribbon at lower temperature ($T_{\mathrm{ref}}$ =40 K). The nodes of angular discretization and spectral discretization are the same as those in the case of $T_{\mathrm{ref}}$=300 K. Fig. 4(a) displays the width-dependent dimensionless thermal conductivity defined as

$$\psi = \frac{\Lambda \int_0^W q_x \mathrm{d}y}{-\lambda_{\mathrm{bulk}} \dfrac{\mathrm{d}T}{\mathrm{d}x} W^2} \tag{46}$$

with the average phonon mean-free-path $\Lambda = \sum_p \int C_{\mathbf{k},p} \mathbf{v}_{\mathbf{k},p} \tau_{\mathrm{C}} \mathrm{d}\mathbf{k} \Big/ \sum_p \int C_{\mathbf{k},p} \mathrm{d}\mathbf{k}$. The physical meaning of $\psi$ is the ratio of average heat flux to the heat flux in ballistic limit. When the ribbon width is relatively small, phonon momentum is mainly dissipated by the boundary and thus the dimensionless thermal conductivity decreases with increasing width as the rarefaction effect decreases. As the normal scattering rate increases, the dimensionless thermal conductivity exhibits a slower decrease and reaches a minimum (the so-called phonon Knudsen minimum [17,22]) at about $W$=2 μm. The phonon Knudsen minimum represents a transition from ballistic transport regime to hydrodynamic transport regime. When the normal scattering rate predominates in thermal transport, the dimensionless thermal conductivity starts to increase with the ribbon width and reaches a maximum at about $W$= 30 μm, which shows clear evidence of hydrodynamic phonon transport. With the ribbon width further increasing, the dimensionless thermal conductivity decreases rapidly with width, implying the significant effect of umklapp scattering on thermal transport. A more detailed discussion of the dynamics of steady-state hydrodynamic phonon transport in graphene ribbon can be found in Ref. [22]. Fig. 4(b), (c) and (d) show the dimensionless cross-sectional heat flux distribution (normalized by the heat flux in bulk limit) obtained by the present DUGKS and the DOM. For different ribbon widths, the heat flux profiles are almost parabolic, indicating that the thermal transport is in or near the hydrodynamic regime. Furthermore, the present DUGKS produces more accurate solution than the DOM with the same coarse mesh $N_x \times N_y = 5 \times 100$ for all the ribbon widths. This comes from the fact that the PBE is stiff in the hydrodynamic transport regime, where the MFP of phonon normal process is much smaller than the width of the graphene ribbon. As the DOM in Ref. [22] adopts the simple step scheme without considering the phonon scattering scale in the evaluation of interface numerical flux, an accurate capture of the phonon dynamics requires the spatial mesh size smaller than the minimum phonon MFP. Thus the DOM has to use a very dense mesh to obtain grid-independent results, which is very computationally expensive. In contrast, a spatial mesh size larger than the phonon MFP can be used in the asymptotic-conserving DUGKS due to the inclusion of scattering process in the evaluation of interface numerical flux. Therefore, the present DUGKS holds advantage over the traditional DOM in terms of numerical accuracy and efficiency in or near the hydrodynamic transport regime. This logic also applies to the simulation of second sound in graphene ribbon, which requires the relaxation time of normal scattering process much smaller than the characteristic time of external heat pulse and makes the solution of PBE stiff. In this situation, the present DUGKS becomes a unique available tool to resolve the transient hydrodynamic phonon transport, as to be discussed in Section 4.

## 4. Results and Discussion

After a solid validation in Section 3, the DUGKS scheme for PBE under Callaway's dual relaxation model is applied for a direct numerical simulation of second sound propagation in graphene ribbon in this section. We consider two cases which are the paradigm process of two available experimental techniques for the detection of second sound: the heat pulse propagation through a graphene ribbon in Section 4.1 and transient thermal grating in graphene ribbon in Section 4.2.

### 4.1 Heat pulse propagation through a graphene ribbon

Heat-pulse techniques have been widely used in the past to detect second sound in three dimensional bulk materials [13,14]. The paradigm schematic of the heat pulse experiment is displayed in Fig. 5. This experiment is started with a short-duration heat pulse exerted on one side of a sample. The temporal temperature response on the other side of the sample will be monitored by a detector. If heat travels across the sample as second sound or ballistic pulse, a wavelike peak will be observed in the backside detector. Although the window condition of second sound in graphene has been formulated by theoretical prediction [16], a direct simulation of second sound propagation in graphene ribbon in such kind of heat pulse experiment has never been achieved due to the lack of methodology as explained in previous part of this article. We intend to accomplish such a direct modeling as guidance for the future experimental detection of second sound in graphene based on our developed numerical scheme. Firstly an infinitely wide graphene ribbon with the length of $L$ is considered to uncover the general features of second sound propagation in Section 4.1.1 and Section 4.1.2. The impact of the lateral boundary in a finite-width graphene ribbon on the dynamics of second sound will be discussed in Section 4.1.3 later. The boundaries of the ribbon are assumed to be adiabatic and treated by the diffuse scheme. At $t$=0, a single square heat pulse with an amplitude $q_0$ and a duration $t_p$ is input on the left boundary of the ribbon ($x$=0). The power of the heat pulse is small enough to ensure a small temperature rise above the ambient temperature. The final temperature of this isolated system after reaching steady-state is determined as $T_{final}=T_{ref}+q_0 t_p/(C_V L)$ and is fixed in all the later simulations. The thermal transport process is modeled by the proposed DUGKS with a spatial grid $N_x$=200, spectral nodes $N_{LA}=N_{TA}$=20 and $N_{ZA}$=100, and angular resolution $N_\theta$=512 after an independence check. The CFL number is set to be 0.852.

4.1.1 Dynamics of second sound propagation

Since hydrodynamic phonon transport is likely to take place in a crystal of substantially higher purity [22], we firstly consider the propagation of heat pulse with $t_p$ =1 ns and $q_0=10^9$ W/m$^2$ through an isotopically pure graphene ribbon with $L$=500 μm around $T_{ref}$=30 K. The length of the graphene ribbon and the ambient temperature are chosen to lie within the window condition of second sound of graphene [16]. Note the graphene ribbon size and system temperature were outside the window condition in a few previous attempts to simulate second sound propagation by non-equilibrium molecular dynamics [42,43] due to too small sample size and too high temperature. The temporal backside temperature response of the graphene ribbon predicted by the present simulation is shown in Fig. 6(a), where several peaks (at time instants $t_1$, $t_2$, …, $t_7$ separately) are obtained clearly. The nature of the temperature peaks can be deduced from both the peak shape and the arrival time. The first two peaks have a very narrow and sharp shape, which is the feature of ballistic heat pulse [11-14]. This can be further confirmed by the arrival times of them: $t_1$=24.145 ns and $t_2$=37.85 ns, which gives the speed of the heat pulse as $v_1=L/(t_1-t_p)=2.132\times10^4$ m/s and $v_2=L/(t_2-t_p)=1.357\times10^4$ m/s very close to the phonon group velocity of LA mode and TA mode ($v_{g,LA}$=2.13×10$^4$ m/s, $v_{g,TA}$=1.36×10$^4$ m/s) respectively. Thus the first two temperature peaks are LA and TA ballistic heat pulses respectively. The third peak arrives at $t_3$=76.55 ns and has a similar shape with the LA and TA ballistic pulse yet with a very small amplitude. This peak is mainly attributed to the reflection signal of LA ballistic pulse from the left edge of graphene ribbon. Different from the former three peaks, the fourth peak has a smoother and broader shape with its arrival time $t_4$=236.15 ns, corresponding to a speed, $v_4=L/(t_4-t_p)=2.122\times10^3$ m/s, much smaller than that of the ballistic pulses. These features indicate that the fourth peak is the second sound, which will be further confirmed by the heat pulse speed later. Second sound takes place in the hydrodynamic transport regime where the normal phonon scattering is dominant and the crystal momentum is almost conserved. The energy dissipation due to minor resistive phonon scattering accounts for the broadening of the second sound peak. In contrast to the strong non-equilibrium nature of ballistic pulse due to rare phonon-phonon scattering, the second sound is a near-equilibrium process due to sufficient phonon normal scattering and thus shows a smoother shape. The rest peaks have similar shape as that of the second sound pulse (fourth peak) and are essentially the reflection signals of the second sound from the left edge of graphene ribbon. This can be deduced from the arrival time of them: $t_5$=732.95 ns, $t_6$=1231.55 ns and $t_7$=1730.60 ns, which satisfy the relation $t_7-t_6\approx t_6-t_5\approx t_5-t_4\approx 2(t_4-t_p)$. The reflection signal of second sound is much weaker and the first second sound peak is often concerned in the practical experimental measurement.

To declare the transient dynamics of second sound propagation, we display the early temporal evolution of the temperature profile within the ribbon in Fig. 6(b). Fig. 6(b) demonstrates that thermal transport does exhibit wave-like behavior. The input heat pulse starts to divide into two peaks and then into three peaks during the propagation, corresponding to the LA and TA ballistic pulses and the second sound in Fig. 6(a). The ballistic pulse, as the individual free pass of phonons, is excited and formed as soon as the input heat pulse is exerted on the left edge of the graphene ribbon. In comparison, it takes some time for the interaction of phonon modes through normal process to form the second sound as a collective motion. To further uncover the contribution of different phonon polarizations to the several heat pulses during propagation, we displays the total heat flux as well as the heat flux from different phonon polarizations at $t$=20 ns in Fig. 6(c). The LA ballistic pulse is fully contributed by the LA phonon mode, whereas the TA ballistic pulse is mainly contributed by the TA phonon mode in spite of some contribution from LA and ZA phonon modes. The second sound is almost contributed by the ZA phonon modes, with tiny contribution from other phonon modes. In Fig. 6(c), another peak is observed in the heat flux component of ZA mode except the second sound peak. This peak has a speed of propagation comparable to the speed of ballistic pulse and is in fact due to ballistic transport of high-frequency ZA phonons. However, the ZA ballistic pulse has very small amplitude and is thus shaded by the signals of other pulses.

The second sound is usually characterized by a speed $v_{ss}$, which can be calculated in the hydrodynamic limit as [16,21],

$$v_{ss}^{limit} = \frac{\sum_p \int \hbar v_{g,x} k_x \frac{\partial f_R^{eq}}{\partial T} \frac{d\mathbf{k}}{D_0}}{\sqrt{\sum_p \int \hbar \omega \frac{\partial f_R^{eq}}{\partial T} \frac{d\mathbf{k}}{D_0} \cdot \sum_p \int \frac{\hbar k_x^2}{\omega} \frac{\partial f_R^{eq}}{\partial T} \frac{d\mathbf{k}}{D_0}}}. \tag{47}$$

According to Eq. (47), the theoretical value of second sound speed in graphene is $2.012\times10^3$ m/s, slightly different from the speed of the fourth peak in Fig. 6(a). The minor difference can be explained by the fact that the second sound pulse has also tiny contribution from the ballistic TA and LA modes as shown in Fig. 6(c), which makes its propagation speed a bit higher than the theoretical value. To separate the contribution from ballistic modes, the second sound speed shall be evaluated when the ballistic pulse has been dissipated to be negligible. As indicated in Fig. 7, the LA and TA ballistic pulses are dissipated much faster than the second sound after the first arrival on the right side of the graphene ribbon. Thus the second sound speed can be calculated via the time delay between the last two reflection signals in Fig. 6(a) as: $v_{ss}=2L/(t_7-t_6)$ $=2.004\times10^3$ m/s, which is indeed very close to the theoretical value in the hydrodynamic limit. Note heat transport in the present condition is exactly in the hydrodynamic limit, as to be shown quantitatively later. This provides a further quantitative confirmation of second sound signal in our preceded analysis.

4.1.2 Impact of intrinsic resistive scattering

The second sound propagation in graphene is expected to be very sensitive to the ambient temperature as the relative strength of normal scattering and resistive scattering will vary a lot versus temperature. In Fig. 8(a), we present the backside temperature response for heat pulse numerical experiment in the isotopically pure graphene ribbon (the abundance of $^{13}$C $c$=0%) with a length $L$=500 μm at various average temperatures. With the temperature elevating, the wavy profiles and arrival times of both LA and TA ballistic pulses do not change very much, whereas their amplitude are diminishing. This can be explained by the fact that both the group velocities of LA and TA phonons are constant based on the adopted phonon dispersion relation (c.f. Appendix A). In comparison, the scattering rates of them will increase with increasing temperature and thus deteriorate the strength of ballistic transport. The second sound peak will arrive slightly earlier at higher temperature, which is attributed to two competitive effects. On one hand, more high-frequency ZA phonons with larger group velocity will be excited at higher temperature and make the average speed thus the second sound speed larger, as given by the theoretical value predicted by Eq. (47) in the hydrodynamic limit shown in Fig. 8(b). On the other hand, more frequent umklapp scattering at higher temperature will induce resistance to second sound and reduce its speed comparing to the value in the hydrodynamic limit, as shown in Fig. 8(b). The first effect dominates over the second effect thus resulting in an overall increase of second sound speed with increasing temperature. The speed reduction effect of umklapp scattering is also quantified by the second sound speed (normalized by the value in hydrodynamic limit) with respect to the ratio of the average linewidth of intrinsic resistive scattering to that of normal scattering (defined as $\Gamma_{\mathrm{R}}/\Gamma_{\mathrm{N}}=\sum_p\int C_{\mathbf{k},p}/\tau_{\mathrm{R}}\frac{d\mathbf{k}}{D_0}\Big/\sum_p\int C_{\mathbf{k},p}/\tau_{\mathrm{N}}\frac{d\mathbf{k}}{D_0}$), as is also displayed in Fig. 8(b). In general, the normalized second sound speed decreases with increasing $\Gamma_R/\Gamma_N$, which increases at elevating temperature. Especially, the second sound speed approaches the value in hydrodynamic limit when the value of $\Gamma_R/\Gamma_N$ goes to zero (e.g. $T_{ref}$=30K). It is also observed in Fig. 8(a) that the second sound peak generally has smaller amplitude at elevating temperature and finally vanishes at about $T_{ref}$=50 K. This is caused by the faster exponentially increasing umklapp scattering rate than the cubically increasing normal scattering rate with respect to temperature based on the adopted phonon relaxation time expressions (c.f. Appendix A). As the umklapp processes gradually become dominant over normal process, phonon momentum and thus the collective behavior of phonons is largely destructed. Therefore, the formation of second sound will be destroyed and heat transport transmits to a diffusive like regime. To further demonstrate this point, Fig. 8(c) shows the temporal evolution of temperature profile within the graphene ribbon at $T_{ref}$=100 K. In comparison to the result in Fig. 7(b), the result in Fig. 8(c) resembles the solution of Fourier's heat diffusion equation, which predicts always a highest temperature at the pulse heating side of the graphene ribbon. Based on the present simulation in graphene ribbon with length from $10^{-6}$ m to $10^{-3}$ m at ambient temperature from 30 K to 100 K, we provide a second sound window contour as a function of temperature and ribbon size as shown in Fig. 8(d). This contour is determined based on a dimensionless temperature defined as $\Theta=(T_{max}-T_{final})/T_{final}$ with $T_{max}$ denoting the maximum value in temperature response curve after all the ballistic pulses. The crucial distinguishing feature of second sound from ballistic transport and diffusive transport is that the temperature $T_{max}$ is higher than the final temperature $T_{final}$. In other words, the region where $\Theta>0$ in this contour represents the window condition where second sound is possible to take place. For instance, in a graphene ribbon with length $L$=100 μm, the second sound window lies approximately between 30~60 K. At a fixed ribbon size, the second sound will take place at neither too high nor too low temperature, since too high temperature will induce resistive scattering and thus diffusive transport whereas too low temperature will foster ballistic transport. At a fixed temperature, the second sound will take place at neither too large ribbon nor too small ribbon due to the similar reason.

The effect of carbon isotopes $^{13}C$ on the second sound propagation is then studied in a graphene ribbon with the length $L$=500 μm at a fixed ambient temperature $T_{ref}$=30 K. With increasing isotope abundance, the second sound peak gradually diminishes and finally disappears when the abundance of $^{13}C$ is larger than about 3.0%, as shown in Fig. 9(a). This is caused by the increasing isotopic scattering, which degrades the momentum of phonons and hydrodynamic effect. On the other hand, the second sound peak arrives at a later instant with the isotope abundance increasing, as shown in Fig. 9(a). The weakening effect and retarding effect of isotope on second sound is further quantified by the dimensionless temperature $\Theta=(T_{max}-T_{final})/T_{final}$ and the arrival time versus the isotope abundance in Fig. 9(b). Generally, the dimensionless temperature characterizing the strength of second sound decreases with increasing isotope abundance, whereas the arrival time shows an increasing trend instead. At very small isotope abundance, the even decrease of second sound strength is caused by stronger ballistic transport in isotopically purified sample. The decrease of arrival time at higher isotope abundance infers that the second sound speed in graphene is dependent on the isotope resistive scattering, which is not taken into account in the Eq. (47) as the hydrodynamic limit. When the resistive scattering (both umklapp scattering and isotope scattering) is taken into account, the second sound will propagate with a speed slower than the value in hydrodynamic limit as calculated with only normal scattering considered. Furthermore, the decrease of second sound speed due to isotope scattering can be approximately fitted by an empirical linear relation as $v_{ss} = v_{ss}^{limit}(1-\eta\Gamma_R/\Gamma_N)$ (c.f. similar expressions for second sound in three-dimensional solids in Refs. [44,45]) as shown in Fig. 9(c), where $\eta$ ranges from 1.894 to 2.489 depending on the ambient temperature. The second sound speed will drop dramatically against the linear law when the dimensionless temperature $\Theta$ approaches to zero, where the heat transport actually starts to transmit from the hydrodynamic regime to the diffusive regime. Within the window condition, the reduction of second sound speed due to the intrinsic resistive scattering is appreciable and at most about 20 percentages. This effect is often neglected in previous theoretical discussions of second sound in graphitic materials [16,21] and shall be kept in mind in the future experimental detection.

4.1.3 Impact of extrinsic resistive scattering

In the preceded subsections, the graphene ribbon is assumed to be infinitely wide. In order to be closer to the actual conditions, we consider a graphene ribbon with finite length and width, as shown in Fig. 5(b) and study the impact of lateral boundary scattering (extrinsic resistive scattering) on the second sound propagation. The lateral edges of the ribbon are assumed to be adiabatic and treated by diffuse scheme. Other conditions are the same as those in the infinitely wide case. Fig. 10(a) shows the backside temporal temperature response curves at various ribbon widths. The temperature response at a very large width $W$=300 μm is almost consistent with that in the infinitely wide case. With the ribbon width decreasing, the peaks of both ballistic pulses and second sound decrease until the vanishing of the second sound peak at $W$=20 μm. On the other hand, the lateral edge will also reduce the speed of second sound as the arrival time increases with decreasing ribbon width. For ribbon width much larger than the MFP of the dominant normal processes, most of the phonons (mainly ZA phonons) contributing to second sound will encounter normal scattering among themselves with few phonons impacted by the lateral boundary, as shown in Fig. 10(b). In this situation, the width size actually lies within the window condition of phonon Poiseuille flow, where the second sound can be well formed and propagated through the graphene ribbon. However, when the ribbon width becomes comparable to or even smaller than the MFP of normal processes, the ZA phonons will be much impacted by the lateral boundary except few phonons propagating along the length direction, which will inhibit sufficient phonon-phonon normal scattering and thus deteriorate the formation as well as the propagation of second sound. Therefore, to observe second sound in experimental measurement, we have to use a graphene ribbon sample with sufficiently large width (>20 μm).

**4.2 Transient thermal grating in a graphene ribbon**

Besides the common heat pulse technique, the transient thermal grating (TTG) is another experimental technique for the detection of second sound. Due to the contactless and non-destructive characterization and the inherently high absolute accuracy, the TTG spectroscopy has been widely used to measure the effective thermal conductivity of nanostructures [46-48] and recently used to detect the second sound in graphite [19]. In this experiment, the sample is impulsively heated by two crossed laser beam, resulting in a sinusoidal periodic temperature profile. Another laser is used to probe the transient thermal decay by measuring the intensity of light diffracted from the sample surface. The profile of thermal decay curve depends on the nature of thermal transport in the sample. Taking one-dimensional TTG as an example, the initial temperature in the sample is given mathematically as $T(x,t=0)=T_{\mathrm{ref}}+\Delta T_{\max}\cos(qx)$ where $x$ is the spatial coordinate, $t$ is the time, $\Delta T_{\max}$ is the amplitude of temperature distribution and $q$ is the spatial wave vector, related to the grating period $L$ through $q=2\pi/L$. In the diffusive regime where the thermal grating period $L$ is much larger than the mean free paths of phonons, the classical Fourier's heat diffusion theory yields an exponential thermal decay of the form $T(x,t)=T_{\mathrm{ref}}+\Delta T_{\max}\cos(qx)\exp(-\lambda_{\mathrm{bulk}}q^2t/C_V)$ with $\lambda_{\mathrm{bulk}}$ and $C_V$ denoting the bulk thermal conductivity and volumetric specific heat, respectively. At sufficiently small grating periods or in the hydrodynamic transport regime, the phonon transport becomes no longer diffusive, leading to non-exponential thermal decay. In this sub-section, we will study the second sound propagation in isotopically pure graphene through a numerical simulation of 1D TTG experiment by the DUGKS scheme for Callaway's dual relaxation model. This case is modeled with $N_x=200$, $N_\theta=128$, $N_{\mathrm{LA}}=N_{\mathrm{TA}}=20$ and $N_{\mathrm{ZA}}=100$ after independence check. The CFL number here is set to be 0.8. We set $\Delta T_{\max}=1$ K in the simulations such that the small temperature difference assumption is satisfied. Since there are no any geometry boundaries in this 1D case, the periodic boundary condition is imposed along the transport direction. For simplicity, we assume the initial state of the whole system to be at local equilibrium given by the Bose-Einstein distribution. Strictly speaking, the initial state has a non-negligible impact on the dynamic process of thermal transport. Nevertheless, this is a reasonable assumption as long as the laser excitation is weak and will not influence the observation of second sound.

Fig. 11(a) shows the numerical results of thermal decay function, defined as $\gamma(t)=(T_A-T\mathrm{ref})/\Delta T_{\max}$, for different grating periods at a fixed ambient temperature $T_{\mathrm{ref}}=30$ K, where $T_A$ denotes the amplitude of the temperature profile here. The thermal decay function displays a typical exponential decay at a very large grating period $L=100$ mm, which indicates that thermal transport is diffusive or quasi-diffusive in this situation. At smaller grating period (i.e. $L=1$ μm, 10 μm, 500 μm), the decay curve of local temperature amplitude becomes strongly non-exponential and demonstrates an oscillating behavior. These oscillating behaviors are attributed to wave-like thermal transport: ballistic heat pulse or second sound. For comparison, the result in the ballistic limit is also included in Fig. 11(a). The thermal decay curve at small grating period ($L=1$ μm) exhibits the same oscillatory feature as that in the ballistic limit. With the grating period increasing, ballistic oscillation gradually become less obvious and another distinguished oscillation with larger amplitude and longer period appears which is attributed to the second sound resonance. In order to further analyze these oscillating curve, Fig. 11(b) shows the Fourier amplitude of the local temperature amplitude as a function of propagation speed of the temperature wave (defined as a product of oscillating frequency and grating period). At small grating period (e.g. $L$=1 μm, 10 μm), the corresponding frequency spectra exhibits three dominant frequency peaks, indicating three different oscillating waves. By comparing the propagation speed and the group velocity of different phonon polarizations, it can be deduced that the two frequency peaks in the inset of Fig. 11(b) are due to TA and LA mode, respectively. As the group velocity of ZA phonon modes largely depends on phonon frequency, the rest frequency peak at a given grating period is attributed to ZA mode. When the second sound resonance takes place at large grating period, only one dominant frequency peak due to ZA mode is obtained. The second sound speed is evaluated to be $2.004\times10^3$ m/s, which agrees well with the value obtained in heat pulse numerical experiment in Section 4.1. Since the effects of ambient temperature and isotope abundance on the dynamics of second sound have been well investigated in the heat pulse numerical experiment, we don't repeat for the present case anymore.

## 5. Conclusion

In this work, we investigate the second sound propagation in graphene ribbon by developing a discrete unified gas kinetic scheme (DUGKS) to solve the space- and time-dependent phonon Boltzmann equation under Callaway's dual relaxation model. The DUGKS is computationally efficient and accurate in the stiff hydrodynamic transport regime, as it retrieves the requirement of time and spatial step smaller than the relaxation time and mean free path of phonons through the inclusion of scattering in the evaluation of interface numerical flux. The transient dynamics of both the heat pulse experiment and transient thermal grating experiment are well reproduced by the DUGKS solution. In the heat pulse numerical experiment, we obtain successively the ballistic LA and TA pulses, the second sound signal mainly contributed by ZA modes in the backside temperature response. The second sound speed is reduced from the theoretical value in the hydrodynamic limit due to finite umklapp and isotope resistive scattering at elevating temperature and isotope abundance. Such a reduction can be at most 20 percentages even within the window condition of second sound. The transverse edge of graphene ribbon will also reduce both the amplitude and speed of second sound because of the boundary resistive scattering. In the transient thermal grating numerical experiment, the entire transition from ballistic to hydrodynamic and to diffusive transport is well captured, with the extracted second sound speed well consistent with that in heat pulse numerical experiment. The present work promotes both the methodology development for solving the phonon Boltzmann equation and an intuitive understanding of the transient hydrodynamic phonon transport. The obtained results will also guide the experimental detection of second sound in graphene in the near feature. Once supplemented with the *ab initio* phonon properties as an input, the present DUGKS framework will provide a more exact resolution of the transient heat transport in graphene as well as in other low-dimensional and three-dimensional materials.

## Acknowledgments

This work has been supported by the National Natural Science Foundation of China (Grant No. 51776054).

## Appendix A. Phonon dispersion and scattering rates of graphene

### 1. Phonon dispersion relation

The primitive unit cell of graphene lattice contains two carbon atoms, resulting in the formation of three acoustic phonon branches and three optical phonon branches over the first Brillouin zone. The optical phonon modes are not taken into account due to their negligible contribution to heat transport because of small group velocities. The three acoustic phonon polarization branches are longitudinal acoustic (LA), transverse acoustic (TA), and flexural acoustic (ZA) phonons. LA modes and TA modes correspond to the in-plane atomic displacements along and perpendicular to the propagation direction of thermal perturbation, whereas ZA modes correspond to the out-of-plane atomic displacements. The frequencies of two in-plane modes (LA and TA) have approximately linear dispersions $\omega=v_{g,LA}k$ and $\omega=v_{g,TA}k$, respectively with the group velocities $v_{g,LA}=2.13\times10^4$ m/s and $v_{g,TA}=1.36\times10^4$ m/s [22]. ZA modes have an approximately quadratic dispersion over a wide range of the 2D Brillouin zone. In order to guarantee the convergence of thermal conductivity in the long-wavelength limit, the renormalized dispersion relation of the ZA mode [39] is employed in this work: $\omega=\alpha_{ZA}[1+(k_c/k)^2]^{1/4}k^2$ with the coefficient $\alpha_{ZA}=6.2\times10^{-7}$ m$^2$/s and the cut-off wave-vector $k_c=0.1k_{max}$ with the maximum wave number $k_{max}=1.5\times10^{10}$ m$^{-1}$.

### 2. Scattering rate

The empirical power-law relaxation time expressions for phonon scattering are considered in the present work as in Ref. [22]. The empirical expression for the umklapp scattering rate is

$$\frac{1}{\tau_U(\omega,p,T)}=\frac{\hbar\gamma_p^2}{\bar{M}\Theta_p v_{g,p}^2}\omega^2 T\exp\left(-\frac{\Theta_p}{3T}\right) \tag{A1}$$

where $\bar{M}$, $\gamma_p$ and $\Theta_p$ denote the average atomic mass, the polarization-dependent Grüneisen parameter and Debye temperature, respectively. For different phonon polarizations of graphene, $\gamma_{LA}=2$, $\gamma_{TA}=2/3$ and $\gamma_{ZA}$=-1.5; and $\Theta_{LA}$=1826.39 K, $\Theta_{TA}$=1126.18 K and $\Theta_{ZA}$=623.62 K. The constituent atom of the ideal graphene unit cell is carbon $^{12}C$; however, naturally occurring graphene will contain the isotope $^{13}C$, which has an abundance of 1.1%. The average mass of an atom in graphene is computed as $\bar{M}=\sum_i f_i M_i = 12+c$ with $M_i$ and $f_i$ the atomic mass and mass fraction of the $i$-th type of isotope, and $c$ being the abundance of $^{13}C$. The empirical expression for the normal scattering rate is given as,

$$\frac{1}{\tau_{\mathrm{N}}(\omega,p,T)}=\left(\frac{k_{\mathrm{B}}}{\hbar}\right)^{b_{\mathrm{N}}}\frac{\hbar\gamma_p^2 V_0^{(a_{\mathrm{N}}+b_{\mathrm{N}}-2)/3}}{\bar{M}v_{\mathrm{g},p}^{a_{\mathrm{N}}+b_{\mathrm{N}}}}\omega^{a_{\mathrm{N}}}T^{b_{\mathrm{N}}} \tag{A2}$$

with $V_0$=8.769634×10$^{-30}$ m$^3$ the average volume of the graphene unit cell and the coefficient pair $a_{\mathrm{N}}$=1, $b_{\mathrm{N}}$=3.

Due to the difference in atomic masses of the constituent carbon isotopes, phonons undergo mass-difference scattering with the heavier isotope atoms. The effect due to the isotopes is incorporated using the mass-difference scattering rate given as,

$$\frac{1}{\tau_{\mathrm{iso}}(\omega,p)}=\begin{cases}\frac{\pi}{2}\Gamma S_0\omega^2 D(\omega,p) & p=\mathrm{ZA}\\ \frac{\pi}{4}\Gamma S_0\omega^2 D(\omega,p) & p=\mathrm{LA,TA}\end{cases} \tag{A3}$$

where the mass difference coefficient is defined as $\Gamma=\sum_i f_i\left(1-M_i/\bar{M}\right)^2 = c(1-c)/(12+c)^2$, $S_0$=2.62×10$^{-20}$ m$^2$ denotes the average area per carbon atom in graphene and $D(\omega, p)$ is the density of states for phonon polarization $p$.

## Appendix B. Explicit second-order upwind scheme for PBE

For simplicity, we just consider the one-dimensional form of PBE. The explicit second-order upwind scheme adopts explicit time integration for both convective term and collision term. The discrete equation is written mathematically as,

$$\begin{aligned}&\frac{g\left(x_i,\mathbf{k}_m,p,t_{n+1}\right)-g\left(x_i,\mathbf{k}_m,p,t_n\right)}{\Delta t}+v_{m,p}^x\frac{g\left(x_{i+1/2},\mathbf{k}_m,p,t_n\right)-g\left(x_{i-1/2},\mathbf{k}_m,p,t_n\right)}{\Delta x}\\&=\frac{g^{\mathrm{eq}}\left(x_i,\mathbf{k}_m,p,t_n\right)-g\left(x_i,\mathbf{k}_m,p,t_n\right)}{\tau_{\mathrm{C}}}\end{aligned} \tag{B1}$$

where the cell interface distribution function is reconstructed as,

$$g\left(x_{i+1/2},\mathbf{k}_m,p,t_n\right)=\begin{cases}g\left(x_i,\mathbf{k}_m,p,t_n\right)+\left(x_{i+1/2}-x_i\right)\sigma\left(x_i,\mathbf{k}_m,p,t_n\right) & v_{m,p}^x>0\\ g\left(x_{i+1},\mathbf{k}_m,p,t_n\right)-\left(x_{i+1}-x_{i+1/2}\right)\sigma\left(x_{i+1},\mathbf{k}_m,p,t_n\right) & \text{else}\end{cases} \tag{B2}$$

The slope σ is calculated by using van Leer limiter methods to avoid unphysical oscillations. With the boundary conditions, it is easy to obtain the solutions for all computational nodes. In order to guarantee the numerical stability, the time step of explicit second-order upwind scheme should satisfy the CFL condition.

## Figure captions

Fig. 1 Flow chart of the implementation of the discrete unified gas kinetic scheme (DUGKS) for phonon Boltzmann equation under Callaway's dual relaxation model. The numbers in the parentheses denote the index of equation in the main text.

Fig. 2 Transient thermal transport in an infinitely wide graphene ribbon with different lengths: (a) dimensionless temperature; (b) dimensionless heat flux. The solid red lines represent the benchmark reference solution based on an explicit second-order upwind scheme[33] with a mesh $N_x$=1000, whereas the dashed lines represent the numerical results by the present discrete unified gas kinetic scheme (DUGKS) scheme with $N_x$=100, $N_{LA}$=$N_{TA}$=$N_{ZA}$=20, $N_\theta$=128.

Fig. 3 Thermal transport in infinitely long graphene ribbon with different widths around $T_{ref}$=300 K: (a) schematic of the heat transport process; (b) cross-sectional distribution of $x$-direction heat flux; (c) width-dependent effective thermal conductivity. The discrete blue circles denote the numerical results by the present discrete unified gas kinetic scheme (DUGKS) with $N_x \times N_y$=5×100, $N_{LA}$=$N_{TA}$=$N_{ZA}$=20 and $N_\theta$=96, the solid red line and cyan dashed line denote the numerical results by discrete-ordinate method (DOM) [22] with a spatial mesh of $N_x \times N_y$=5×1000 and $N_x \times N_y$=5×100 respectively; the discrete black squares represent the *ab initio* Monte Carlo (MC) solution of phonon Boltzmann equation with full scattering term [41]. The inset in (c) shows the effective thermal conductivity in the ballistic limit of small ribbon width.

Fig. 4 Thermal transport in infinitely long graphene ribbon with different widths around $T_{ref}$=40 K. (a) dimensionless thermal conductivity defined in Eq. (46): the discrete blue circles denote the numerical results by the present discrete unified gas kinetic scheme (DUGKS) with $N_x \times N_y$=5×100, $N_{LA}$=$N_{TA}$=$N_{ZA}$=20 and $N_\theta$=96, the solid red line denotes the numerical results by discrete-ordinate method (DOM) with $N_x \times N_y$=5×1000, $N_{LA}$=$N_{TA}$=$N_{ZA}$=20 and $N_\theta$=96. Cross-sectional distribution of $x$-direction heat flux at different widths (b) $W$=10 μm, (c) $W$=30 μm and (d) $W$=100 μm: the discrete blue circles and solid magenta lines denote the numerical results by DUGKS with a spatial mesh of $N_x \times N_y$=5×100 and $N_x \times N_y$=5×400 respectively; the cyan dash line, olive dash-dotted line and black dotted line denote the numerical results by DOM with a spatial mesh of $N_x \times N_y$=5×100, $N_x \times N_y$=5×400 and $N_x \times N_y$=5×1000 respectively.

Fig. 5 Schematic of heat pulse experiment in (a) an infinitely wide graphene ribbon with a length $L$ and (b) a rectangular graphene ribbon with a length $L$ and width $W$.

Fig. 6 Heat pulse in infinitely wide graphene ribbon with the abundance of $^{13}$C $c$=0 (isotopically pure) and the ribbon length $L$=500 μm at $T_{ref}$=30 K: (a) backside temperature response; (b) temperature distribution at various time; (c) the total heat flux and branch heat flux at $t$=20 ns.

Fig. 7 The total heat flux and branch heat flux along the length direction of infinitely wide isotopically pure graphene ribbon with a length $L$=500 μm around $T_{\mathrm{ref}}$=30 K at various time: (a) $q_x$; (b) $q_{\mathrm{x,LA}}$; (c) $q_{x,\mathrm{TA}}$; (d) $q_{x,\mathrm{ZA}}$.

Fig. 8 The temperature effect on the propagation of heat pulses in isotopically pure graphene ribbon with length $L$=500 μm: (a) backside temperature response; (b) second sound speed with respect to ambient temperature (the black solid line denotes the analytical value in hydrodynamic limit without considering the resistive process and the blue solid line with circles denotes the results obtained by the present DUGKS solution of PBE under Callaway's dual relaxation model) and the purple line with squares represents the normalized second sound speed with respect to the ratio of the average linewidth of normal scattering to that of intrinsic resistive scattering; (c) the temperature evolution in the ribbon at $T_{\mathrm{ref}}$=100 K; (d) second sound window for isotopically pure graphene (The color scale corresponds to the dimensionless temperature $\Theta=(T_{\mathrm{max}}-T_{\mathrm{final}})/T_{\mathrm{final}}$. Second sound occurs in the hydrodynamic regime where $\Theta>0$).

Fig. 9 Effect of the abundance of $^{13}$C on second sound effect: (a) backside temperature response at $L$=500 μm, $T_{\mathrm{ref}}$=30 K; (b) the arrival time of second sound peak and strength of hydrodynamic transport as a function of isotopic abundance; (c) normalized second sound speed as a function of the ratio of the average linewidth of intrinsic resistive scattering to that of normal scattering.

Fig. 10 Heat pulse in isotopically pure graphene ribbon with different width $W$ and fixed length $L$=500 μm at $T_{\mathrm{ref}}$=30 K: (a) backside temperature response (b) schematic illustration of the effect of ribbon width on normal scattering ($\Lambda_{\mathrm{N}}$ denotes the mean free path of normal processes).

Fig. 11 Thermal decay in isotopically pure graphene ribbon for various grating period at $T_{\mathrm{ref}}$=30 K: (a) The decay curve of local temperature as a function of normalized time (the reference time $t_{\mathrm{ref}}=L/v_{\mathrm{g}}^{\mathrm{max}}$); (b) the frequency spectra of the thermal decay function by using fast Fourier transformation (FFT).

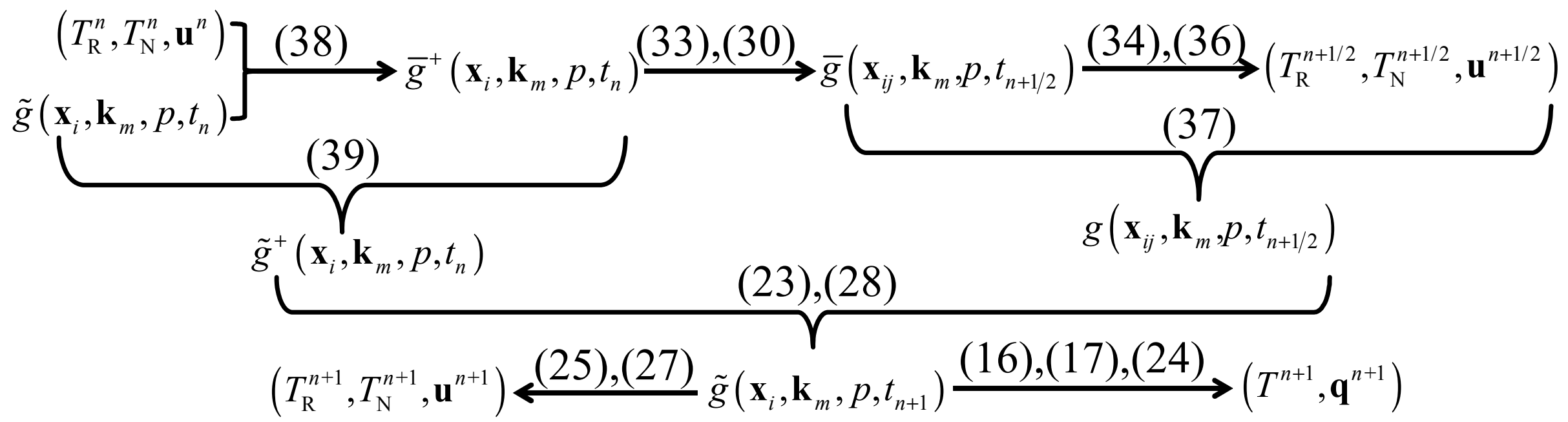

Fig. 1 Flow chart of the implementation of the discrete unified gas kinetic scheme (DUGKS) for phonon Boltzmann equation under Callaway's dual relaxation model. The numbers in the parentheses denote the index of equation in the main text.

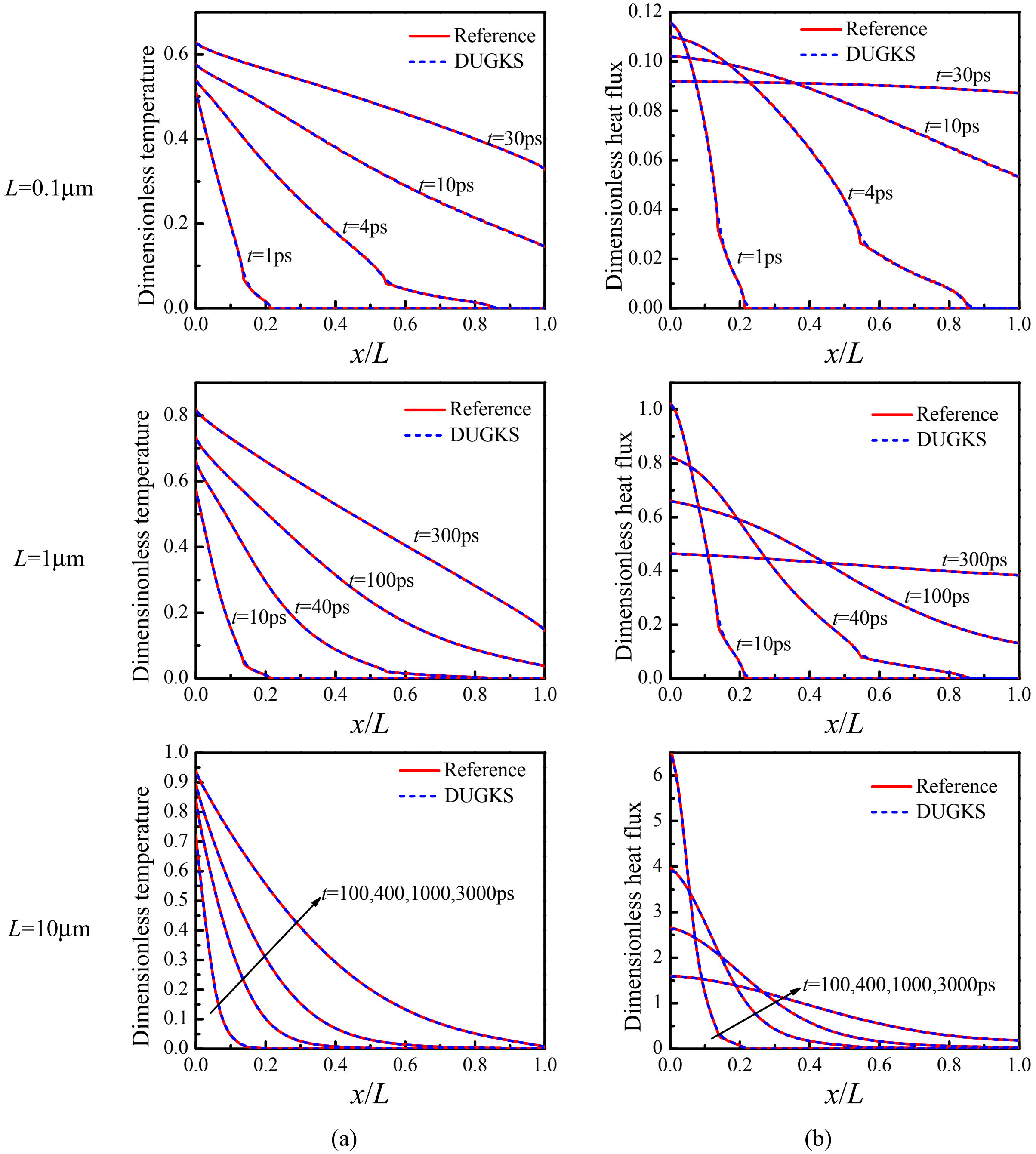

Fig. 2 Transient thermal transport in an infinitely wide graphene ribbon with different lengths: (a) dimensionless temperature; (b) dimensionless heat flux. The solid red lines represent the benchmark reference solution based on an explicit second-order upwind scheme[33] with a mesh $N_x$=1000, whereas the dashed lines represent the numerical results by the present discrete unified gas kinetic scheme (DUGKS) scheme with $N_x$=100, $N_{LA}$=$N_{TA}$=$N_{ZA}$=20, $N_\theta$=128.

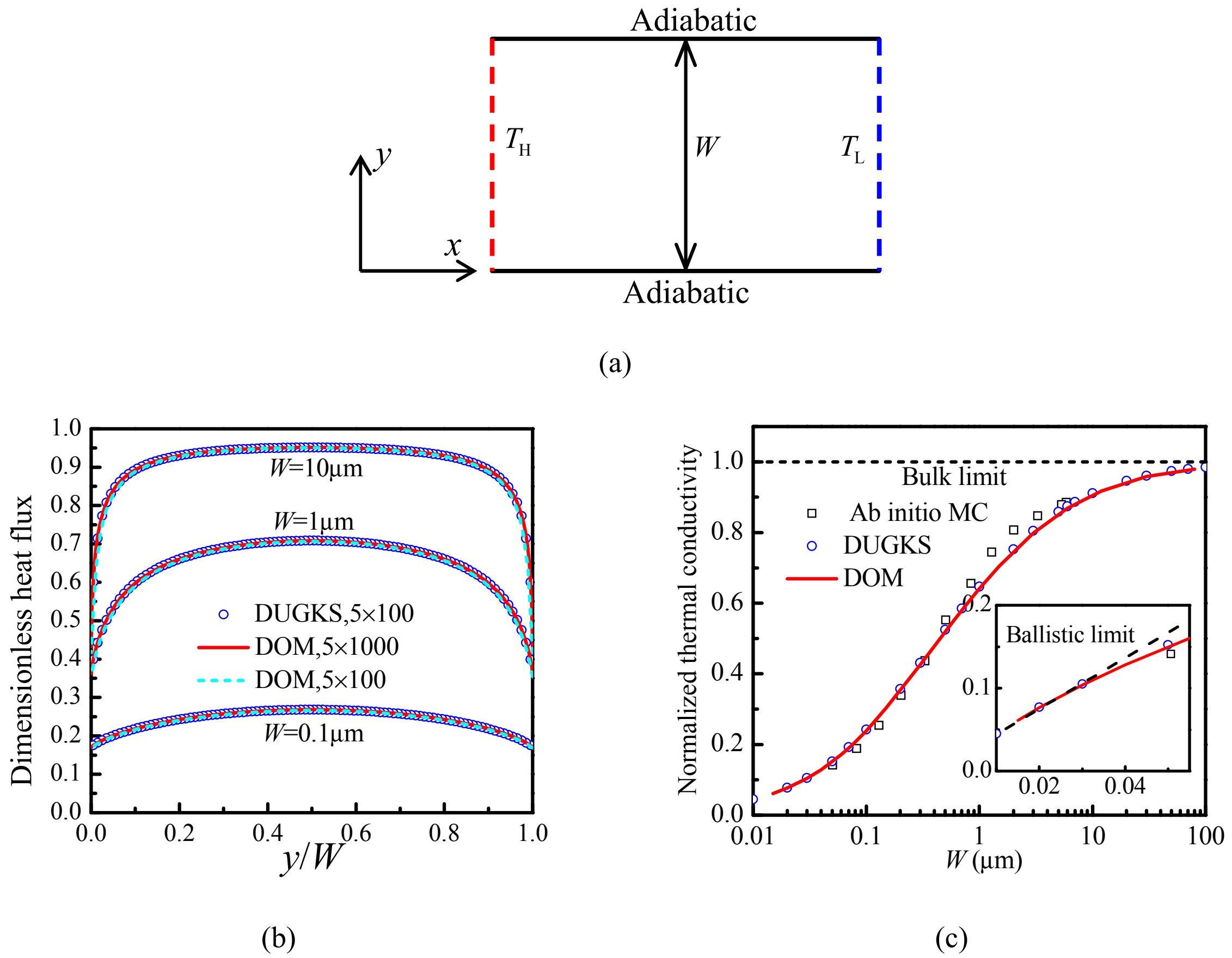

Fig. 3 Thermal transport in infinitely long graphene ribbon with different widths around $T_{\mathrm{ref}}$=300 K: (a) schematic of the heat transport process; (b) cross-sectional distribution of $x$-direction heat flux; (c) width-dependent effective thermal conductivity. The discrete blue circles denote the numerical results by the present discrete unified gas kinetic scheme (DUGKS) with $N_x \times N_y$=5×100, $N_{\mathrm{LA}}=N_{\mathrm{TA}}=N_{\mathrm{ZA}}$=20 and $N_\theta$=96, the solid red line and cyan dashed line denote the numerical results by discrete-ordinate method (DOM) [22] with a spatial mesh of $N_x \times N_y$=5×1000 and $N_x \times N_y$=5×100 respectively; the discrete black squares represent the *ab initio* Monte Carlo (MC) solution of phonon Boltzmann equation with full scattering term [41]. The inset in (c) shows the effective thermal conductivity in the ballistic limit of small ribbon width.

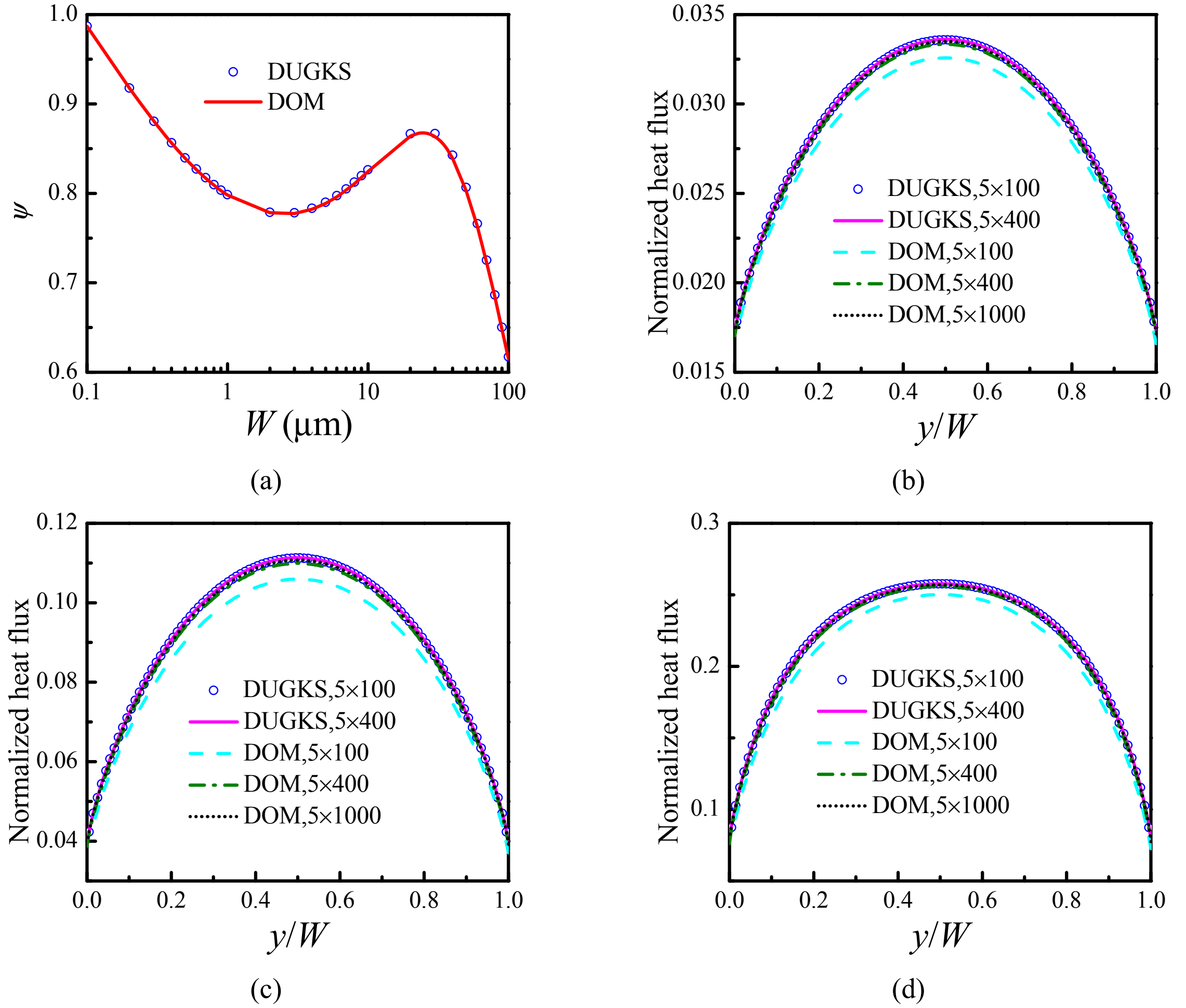

Fig. 4 Thermal transport in infinitely long graphene ribbon with different widths around $T_{\mathrm{ref}}$=40 K. (a) dimensionless thermal conductivity defined in Eq. (46): the discrete blue circles denote the numerical results by the present discrete unified gas kinetic scheme (DUGKS) with $N_x \times N_y$=5×100, $N_{\mathrm{LA}}=N_{\mathrm{TA}}=N_{\mathrm{ZA}}$=20 and $N_\theta$=96, the solid red line denotes the numerical results by discrete-ordinate method (DOM) with $N_x \times N_y$=5×1000, $N_{\mathrm{LA}}=N_{\mathrm{TA}}=N_{\mathrm{ZA}}$=20 and $N_\theta$=96. Cross-sectional distribution of *x*-direction heat flux at different widths (b)$W$=10 μm, (c) $W$=30 μm and (d) $W$=100 μm: the discrete blue circles and solid magenta lines denote the numerical results by DUGKS with a spatial mesh of $N_x \times N_y$=5×100 and $N_x \times N_y$=5×400 respectively; the cyan dash line, olive dash-dotted line and black dotted line denote the numerical results by DOM with a spatial mesh of $N_x \times N_y$=5×100, $N_x \times N_y$=5×400 and $N_x \times N_y$=5×1000 respectively.

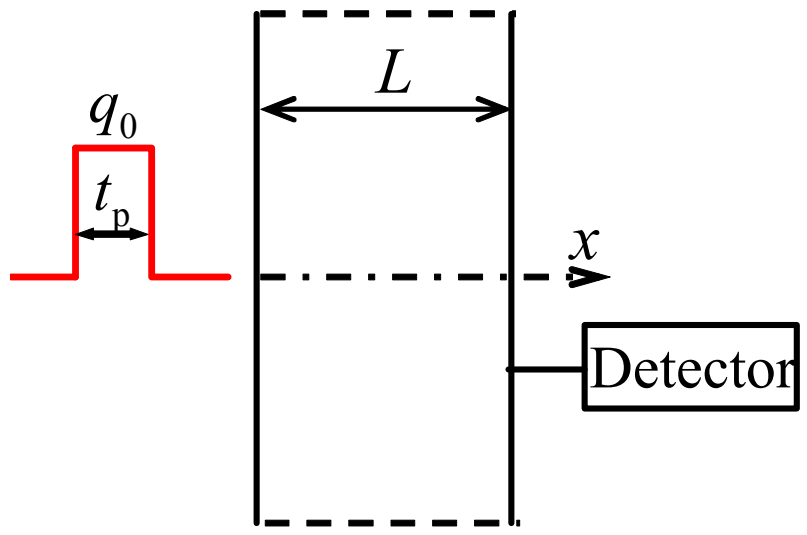

(a)

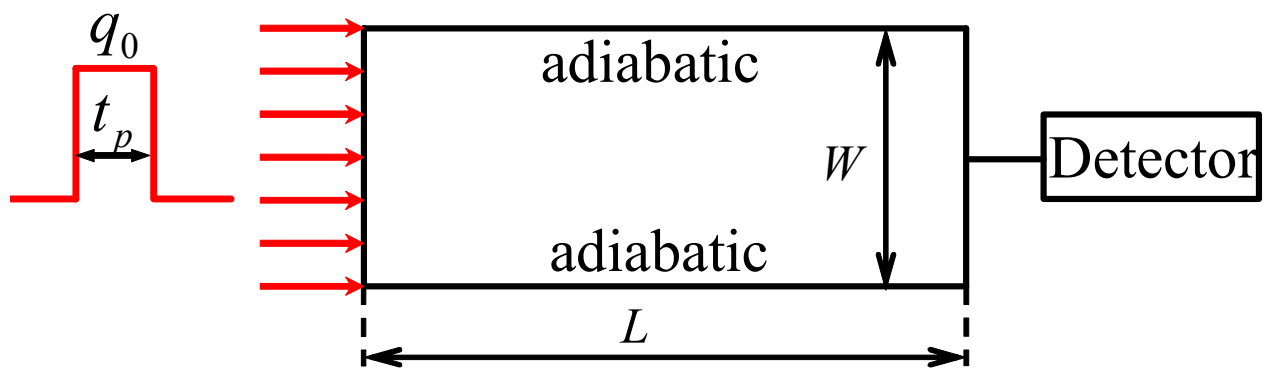

(b)

Fig. 5 Schematic of heat pulse experiment in (a) an infinitely wide graphene ribbon with a length $L$ and (b) a rectangular graphene ribbon with a length $L$ and width $W$.

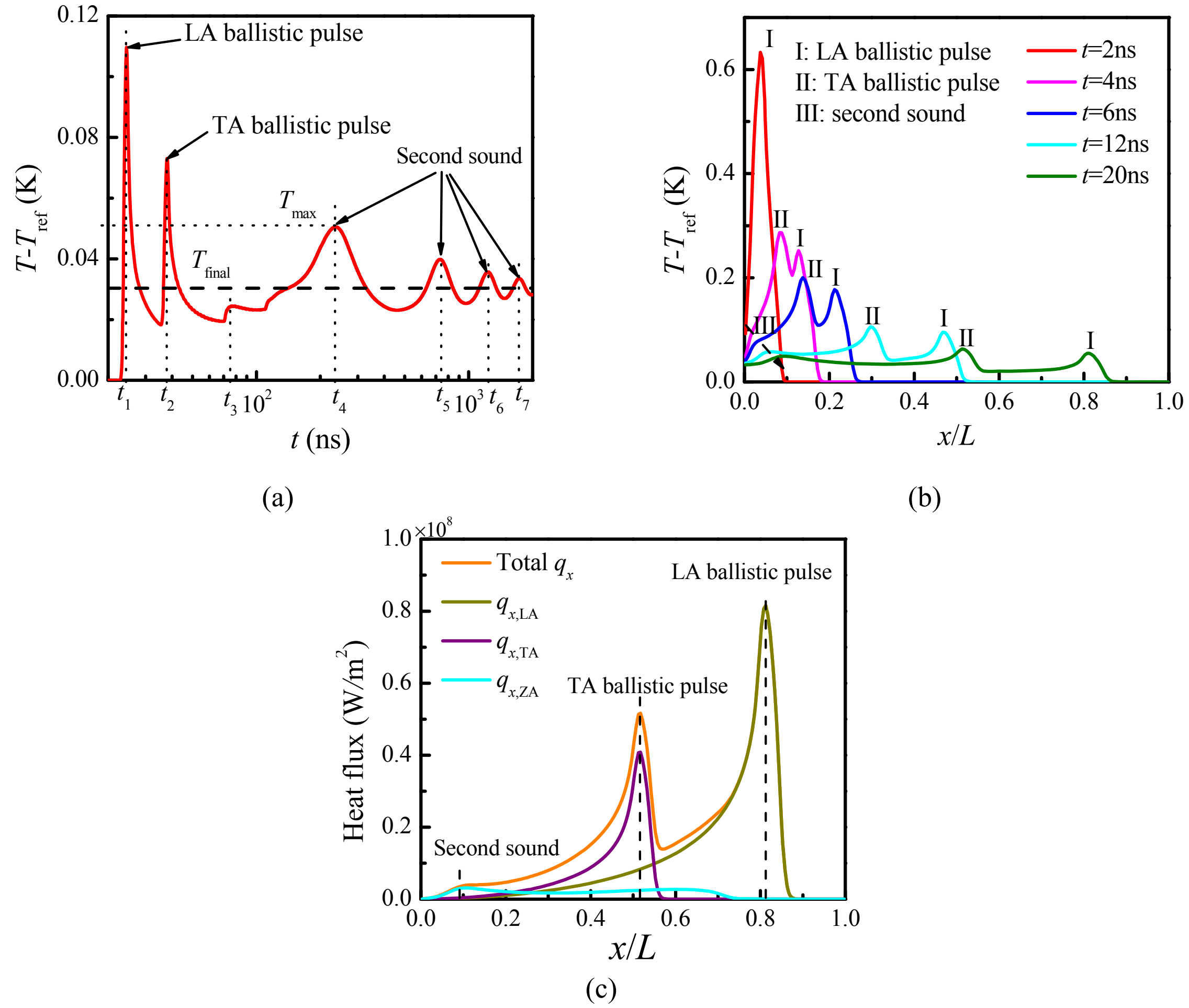

Fig. 6 Heat pulse in infinitely wide graphene ribbon with the abundance of $^{13}$C $c$=0 (isotopically pure) and the ribbon length $L$=500 μm at $T_{\mathrm{ref}}$=30 K: (a) backside temperature response; (b) temperature distribution at various time; (c) the total heat flux and branch heat flux at $t$=20 ns.

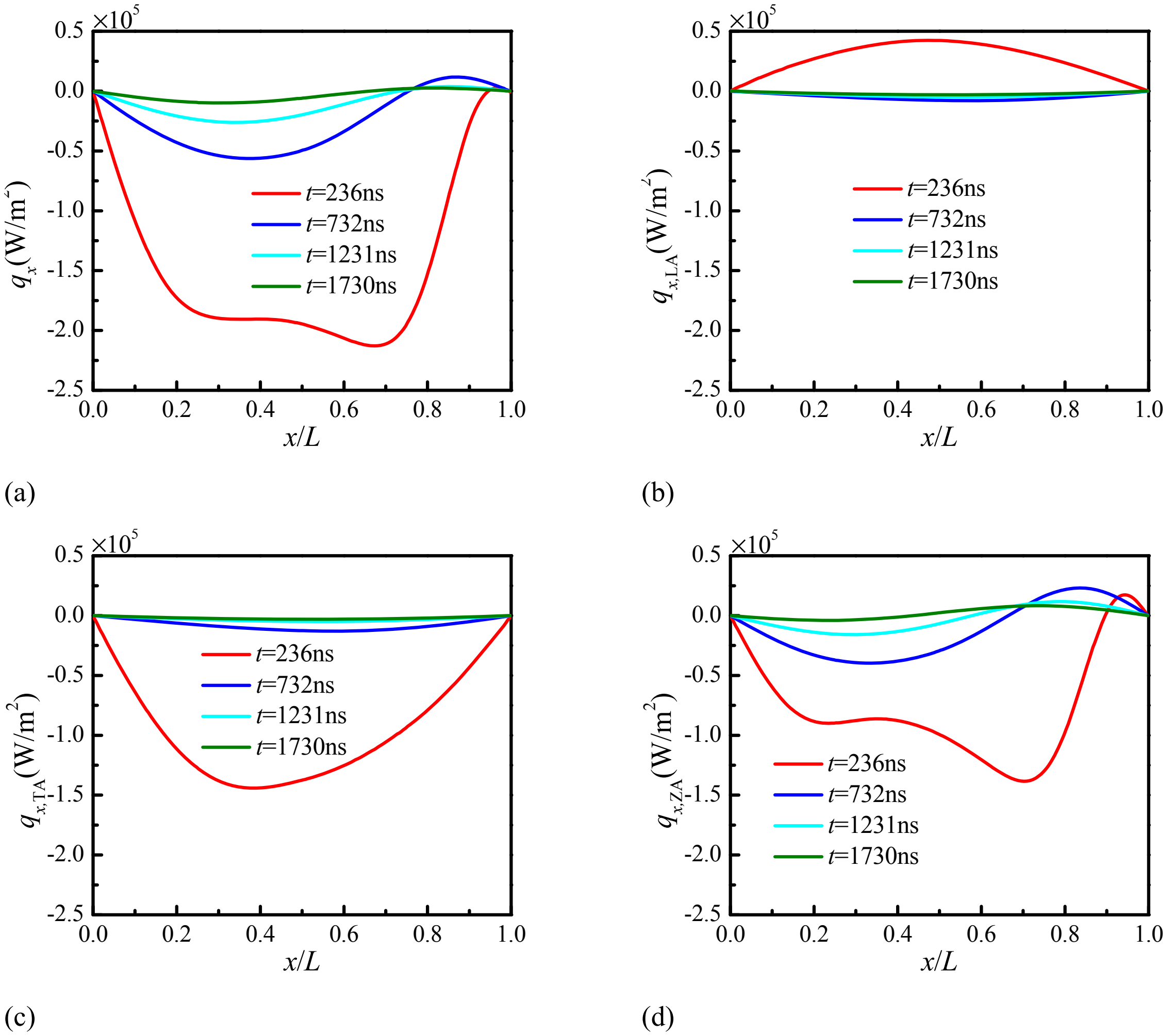

Fig. 7 The total heat flux and branch heat flux along the length direction of infinitely wide isotopically pure graphene ribbon with a length $L$=500 μm around $T_{\mathrm{ref}}$=30 K at various time: (a) $q_x$; (b) $q_{\mathrm{x,LA}}$; (c) $q_{x,\mathrm{TA}}$; (d) $q_{x,\mathrm{ZA}}$.

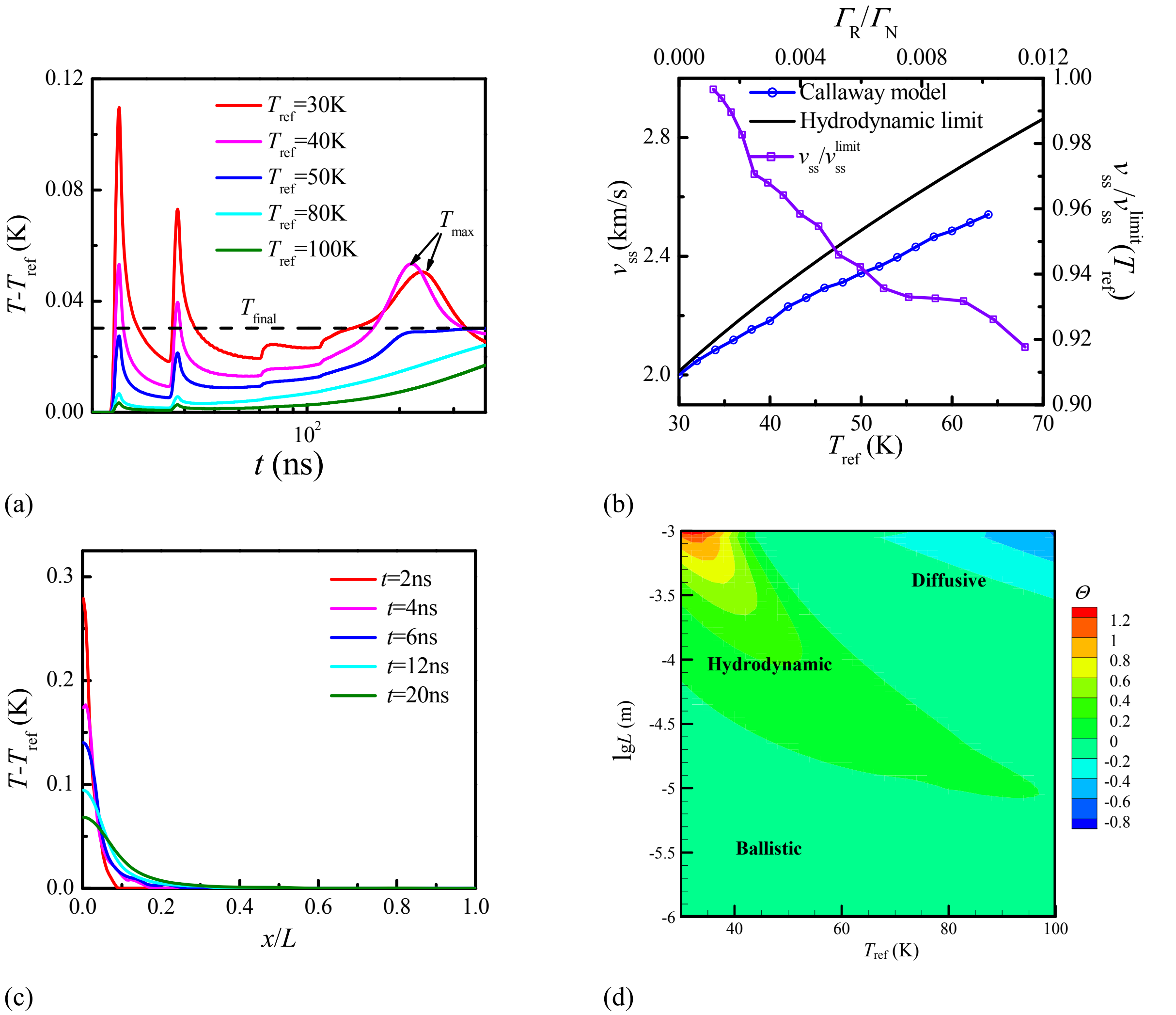

Fig. 8 The temperature effect on the propagation of heat pulses in isotopically pure graphene ribbon with length $L$=500 μm: (a) backside temperature response; (b) second sound speed with respect to ambient temperature (the black solid line denotes the analytical value in hydrodynamic limit without considering the resistive process and the blue solid line with circles denotes the results obtained by the present DUGKS solution of PBE under Callaway's dual relaxation model) and the purple line with squares represents the normalized second sound speed with respect to the ratio of the average linewidth of normal scattering to that of intrinsic resistive scattering; (c) the temperature evolution in the ribbon at $T_{ref}$=100 K; (d) second sound window for isotopically pure graphene (The color scale corresponds to the dimensionless temperature $\Theta=(T_{max}-T_{final})/T_{final}$. Second sound occurs in the hydrodynamic regime where $\Theta>0$).

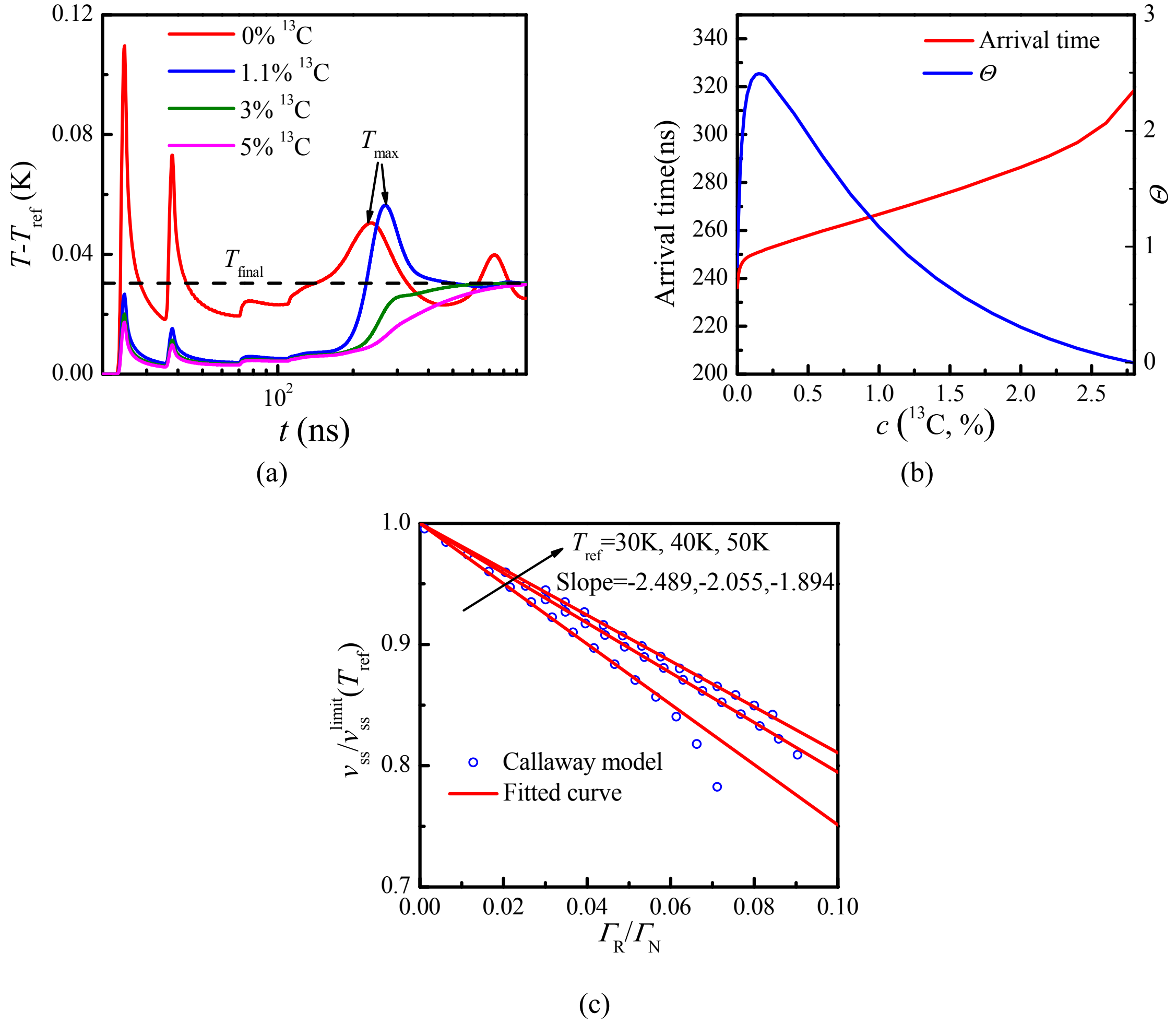

Fig. 9 Effect of the abundance of $^{13}C$ on second sound effect: (a) backside temperature response at $L$=500 μm, $T_{ref}$=30 K; (b) the arrival time of second sound peak and strength of hydrodynamic transport as a function of isotopic abundance; (c) normalized second sound speed as a function of the ratio of the average linewidth of intrinsic resistive scattering to that of normal scattering.

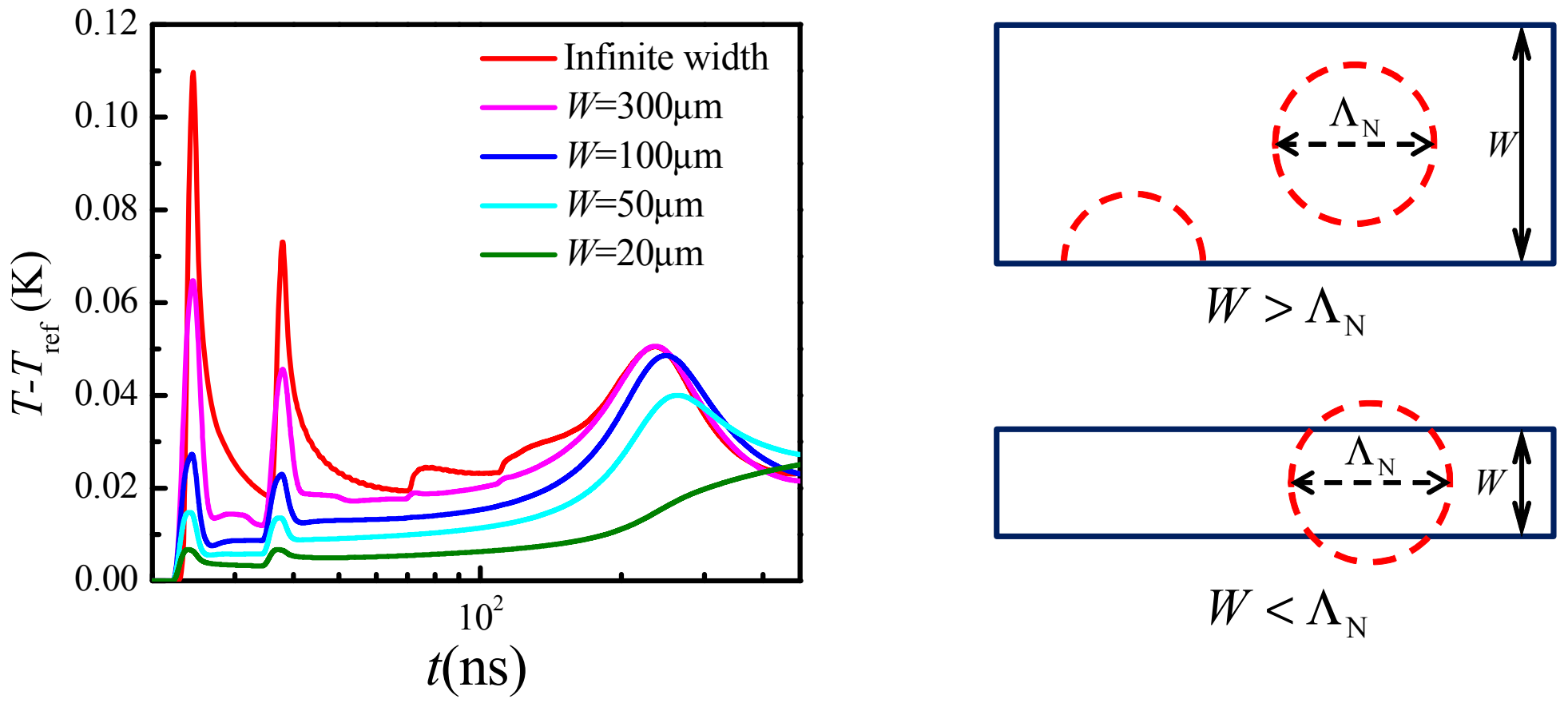

(a) (b)

Fig. 10 Heat pulse in isotopically pure graphene ribbon with different width $W$ and fixed length $L$=500 μm at $T_{ref}$=30 K: (a) backside temperature response (b) schematic illustration of the effect of ribbon width on normal scattering ($\Lambda_N$ denotes the mean free path of normal processes).

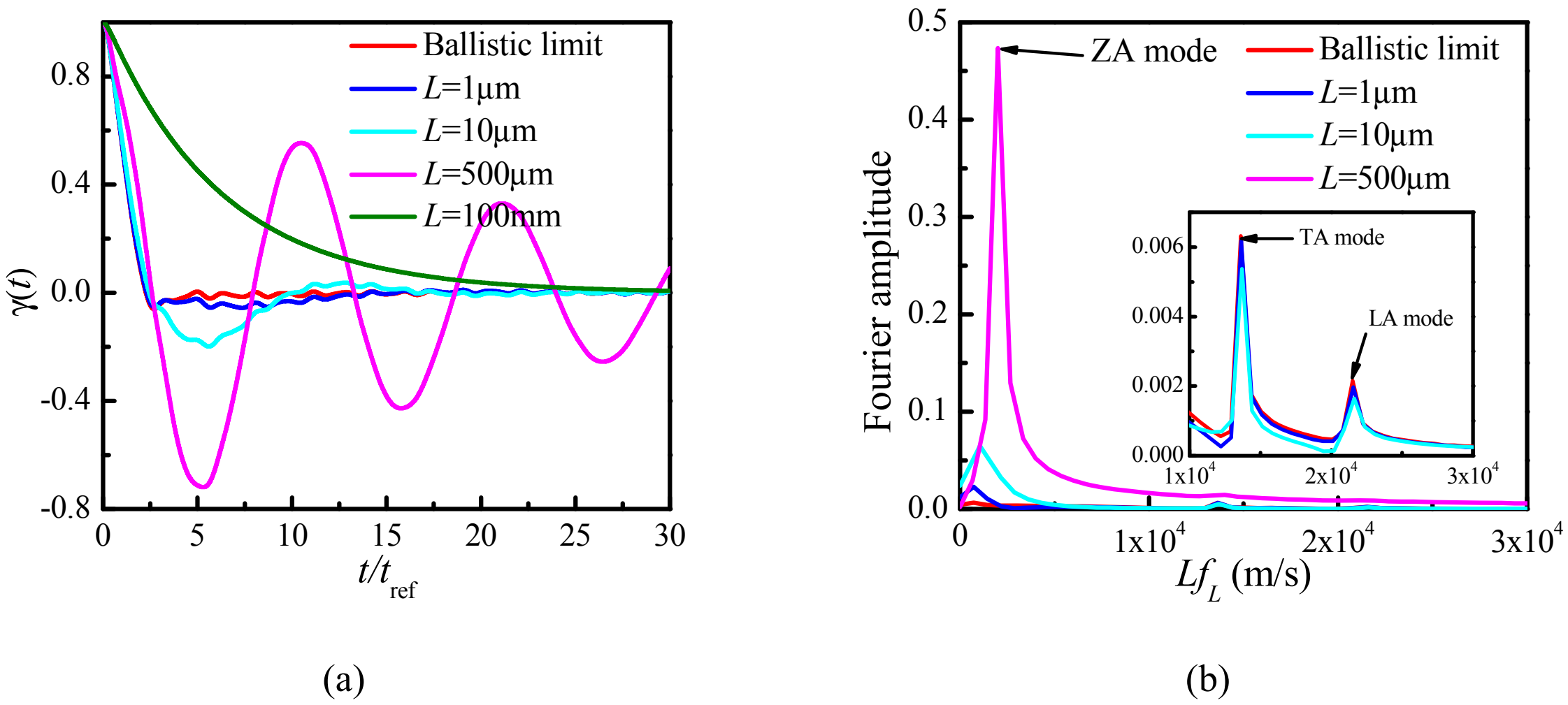

Fig. 11 Thermal decay in isotopically pure graphene ribbon for various grating period at $T_{ref}$=30 K: (a) The decay curve of local temperature as a function of normalized time (the reference time $t_{ref}=L/v_g^{max}$); (b) the frequency spectra of the thermal decay function by using fast Fourier transformation (FFT).